\documentclass[12pt]{article}
\usepackage{amsfonts,amsmath}
\usepackage{amssymb}

\makeatletter \@addtoreset{equation}{section} \makeatother

{\vspace{3mm} }

\def\al{\alpha}

\def\*{\star}

\def\E2{\mathbf{E}}

\newcommand{\be}{\begin{equation}}
\newcommand{\ee}{\end{equation}}
\newcommand{\bee}{\begin{eqnarray}}
\newcommand{\beee}{\begin{array}}
\newcommand{\eee}{\end{eqnarray}}
\newcommand{\eeee}{\end{array}}
\newcommand{\gb}{\beta}
\newcommand{\gga}{\gamma}

\newcommand{\gd}{\delta}

\newcommand{\gk}{\varkappa}
\newcommand{\gep}{\epsilon}

\newcommand{\gs}{\sigma}
\newcommand{\go}{\omega}

\newcommand{\dal}{\dot \alpha}
\newcommand{\dgb}{\dot \beta}
\newcommand{\dgga}{\dot \gamma}

\newcommand{\nn}{\nonumber}
\newcommand{\half}{\frac{1}{2}}

\newcommand{\p}{\partial}

\newcommand{\ff}{\frac}
\newcommand{\dr}{{\rm d}}

\begin{document}
\begin{flushright}
FIAN/TD/08-2021\\
\end{flushright}

\vspace{0.5cm}
\begin{center}
{\large\bf Planar solutions of higher-spin theory I: free field
level}

\vspace{1 cm}

\textbf{V.E.~Didenko, A.V.~Korybut}\\

\vspace{1 cm}

\textbf{}\textbf{}\\
\vspace{0.5cm}  \textit{I.E. Tamm Department of Theoretical
Physics,
Lebedev Physical Institute,}\\
 \textit{ Leninsky prospect 53, 119991, Moscow, Russia }\\
\par\end{center}

\vspace{0.4cm}

\begin{abstract}
\noindent Many black hole solutions of General Relativity are
known to be linearly exact. This opens a way to study them in
gauge theories that apart from gravity contain fields of higher
spin $s>2$. Starting with a black brane in $AdS_4$ we find its
free field higher-spin generalization that respects static and
planar symmetry for all bosonic gauge fields $s\geq 0$. The
solution is found for both the higher-spin curvatures and
potentials in the form suitable for further non-linear analysis
and satisfies the multi copy relation.
\end{abstract}
\newpage
\tableofcontents
\section{Introduction}
Remarkable properties of black holes make them a valuable arena in
different branches of theoretical physics. Particularly, within
the $AdS/CFT$ correspondence \cite{Maldacena:1997re,
Gubser:1998bc, Witten:1998qj} black holes play central role in
description of various thermodynamical phenomena that take place
in the dual $CFT$ theory. Glimpses of deep relation between black
hole entropy and its statistical realization in terms of free
fields have already been available \cite{Klebanov:1996un}, even
before they plunged into gauge/gravity mainstream
\cite{Witten:1998zw}. Lots of investigation is focused on testing
different aspects of the famous correspondence which relates two
theories in the opposite regimes. Given one of the two sides is
strongly coupled, no wonder the duality still remains a
conjecture.

Among many dual pairs there are some that relate simplest CFT
vector models of $N$ free fields on the three dimensional boundary
of $AdS_4$ to a highly non-trivial higher-spin (HS) theory in the
bulk \cite{Klebanov:2002ja}, \cite{Sezgin:2002rt},
\cite{Leigh:2003gk}. Along with graviton the latter contains
interacting scalar and gauge fields of all integer spins. As the
dual $CFT$ is supposed to be free such a duality is of weak-weak
type and therefore is testable at least in perturbations.
Currently this endeavor is mostly confined to tree level since the
conventional HS action principle is still lacking.

Before the $AdS/CFT$ ages, HS theory had its own pace with some
notable milestones which include the free Fronsdal Lagrangian
\cite{Fronsdal}, the first instance of cubic vertices found using
the light-cone approach \cite{BBB}, the observation of relevance
of $AdS$ background for HS interactions \cite{Fradkin:1987ks}, the
full list of cubic interactions classified by Metsaev
\cite{Metsaev:1993}, and eventually the generating Vasiliev system
for any order HS equations \cite{more} (see \cite{Sorokin:2004ie},
\cite{Vasiliev:1999ba}, \cite{Bekaert:2005vh},
\cite{Didenko:2014dwa} for reviews). Higher-order off-shell
formulation that would make quantum analysis accessible is still
not there though there are steps in this direction (see e.g.,
\cite{Boulanger:2011dd}, \cite{Sleight:2016dba},
\cite{Vasiliev:2018zer}, \cite{Misuna:2019ijn}).

It took a while since the original $HS/CFT$ proposal was made
before some checks carried out at the level of three point
functions. In a remarkable feat of technical analysis Giombi and
Yin \cite{GY1} were able to extract certain $3pt$ correlators from
Vasiliev equations of motion and found perfect match with the
boundary expectations. With this work HS theory gained another
boost that largely clarified the status of the duality (for
incomplete list of references see \cite{Maldacena:2011jn},
\cite{Aharony:2011jz}, \cite{Giombi:2011kc},
\cite{Maldacena:2012sf}, \cite{Aharony:2012ns},
\cite{Sleight:2016dba}, \cite{Giombi:2013fka},
\cite{Beccaria:2014xda}, \cite{Vasiliev:2012vf},
\cite{Didenko:2012tv}, \cite{Gelfond:2013xt}).

Little is known however what happens to HS bulk theory in $d+1\geq
3$ at non-zero temperature $T>0$. Typically of gravity theories,
the CFT dual thermal states with $T<T_c$ are associated with
planar black holes that radiate Hawking temperature $T_H$. It
originates from a factorization that removes a deficit angle in
the Euclidian version of a black hole metric thus introducing a
thermal $S^1$. For the boundary theory on a sphere a large
spherical black hole in the bulk corresponds to a thermal state
beyond a critical point $T>T_c$, where a phase transition occurs.

HS theory is very much different as its spin two sector and the
corresponding metric seems to have no invariant meaning under
higher-spin symmetry transformation. This makes the very notion of
horizon debatable. Even in the case of $HS_3/CFT_2$  duality
\cite{Campoleoni:2010zq}, \cite{Henneaux:2010xg},
\cite{Gaberdiel:2010pz} which is much more elaborated due to its
topological nature and the fact that the HS field spectrum can be
made finite, the role of $HS_3$ black hole horizon
\cite{Gutperle:2011kf} (if any) is unclear.

Besides, as Shenker and Yin pointed out in \cite{Shenker:2011zf},
for $d=3$ $U(N)$ vector model on $S^2\times S^1$ the phase
transition takes place at Plankian temperature $T\sim\sqrt{N}$
rather than at $AdS$ scale $T\sim 1$ in sharp contrast  with a
field theory in the adjoint. This fact seemingly indicates the
absence of uncharged spherically symmetric black hole in $AdS_4$
HS theory. This might not be too surprising after all, as
generally in HS theory lower spins source higher spins and $s=2$
is not an exception. Whether it is so with a black brane or not
will be investigated in \cite{DK} at the lowest interaction order.
But general expectation is that a potential HS black hole may have
infinitely many charges. This may result in smearing off the $AdS$
scale phase transition due to infinitely many black hole states.
All in all, the idea of associating thermal states of HS theory to
its exact solutions with proper global symmetry still seems
natural.

The goal of this paper is to set the stage for perturbative
analysis of static HS solutions in four dimensions that in
addition have planar symmetry and to find the associated solutions
at free level. In other words we are interested in HS counterparts
of black branes in perturbations. The problem is considered within
the Vasiliev theory.

One of the reasons for perturbative analysis rather than the all
order one is the HS locality problem that for the time being has
not been yet solved to all orders. To be more specific it is not
clear whether HS vertices for a given set of spins contain finite
number of derivatives beyond cubic order or not
\cite{Bekaert:2015tva}, \cite{Sleight:2017pcz}. This problem is a
daunting challenge within the holographic approach that allows one
at least in principle reconstructing bulk interactions from the
free boundary $CFT$ correlators. This idea was proposed by Petkou
in \cite{Petkou:2003zz} as he anticipated that the procedure of
reconstructing quartic scalar interaction would be straightforward
if perhaps technical. It was solved later on in
\cite{Bekaert:2015tva}. At quartic level one encounters
non-localities that somehow drives the analysis out of control
\cite{Sleight:2017pcz}. Indeed, as was pointed out in
\cite{Sleight:2017pcz}, \cite{Barnich:1993vg}, admitting wild
non-localities in field redefinition trivializes the Noether
procedure and formally reduces interacting theory to the free one.

The state of affairs with $d+1\geq 4$ Vasiliev theory is less
uncertain although the (non)locality problem is a pressing issue
too. The great advantage of the Vasiliev approach is that the HS
equations it reproduces are given in any HS background rather than
on $AdS$. On a practical note this renders the perturbation theory
pretty peculiar being exact in HS 1-form potentials $\go$ while
decomposable in terms of HS Weyl tensors $C$. It turns out in
particular that as shown in \cite{Vasiliev:1989yr} a cubic vertex
(quadratic in $C$) can not be zero once the linear one in $C$ is
present. This implies that there are no field redefinitions that
allows one getting rid of cubic vertex for if there were, the
resulting system would be either formally inconsistent or
contained no free equations. The latter option can not be realized
for cohomological reason. This seemingly leaves no room for any
ambiguity in calculation of boundary observables from the bulk.
Once they are finite they can be checked against $CFT$
expectations. The non-admissible non-locality results into
infinities for these quantities. An urgent problem is therefore to
understand a class of admissible field variables that render
obserbvables finite.

Being formulated in a certain twistor space a natural substitute
for the space-time locality in this approach is a twistor locality
proposed in \cite{Gelfond:2018vmi} and coined the spin-locality
which if present guarantees finiteness of classical theory. In a
series of papers \cite{Gelfond:2018vmi}, \cite{Vasiliev:2016xui},
\cite{Didenko:2018fgx}, \cite{Didenko:2019xzz},
\cite{Gelfond:2019tac}, \cite{Didenko:2020bxd},
\cite{Gelfond:2021two} the spin-locality conjecture was
confirmed\footnote{The observed structure of spin-locality turned
out to have a certain $Z_2$ -- graded form. While spin-local
structures can generate non-local ones in star-product
commutators, in practice they do not for among two of them within
a star-product commutator at least one appears to be spin
ultra-local \cite{Didenko:2018fgx} yielding no non-localities.}
for many non-linear Vasiliev vertices but still those do not yet
cover the full quartic vertex on $AdS$. It is for this reason that
we confined to perturbative analysis in our quest for planar
solutions.

Despite the locality issue, there is a handful of exact solutions
for Vasiliev equations in four dimensions that were constructed
over the years \cite{Sezgin:2005pv}, \cite{Iazeolla:2007wt},
\cite{Didenko:2009td}, \cite{Iazeolla:2011cb},
\cite{Sundell:2016mxc}, \cite{Iazeolla:2017vng},
\cite{Iazeolla:2015tca}, \cite{Iazeolla:2017dxc},
\cite{Aros:2017ror}. Particularly, in \cite{Didenko:2009td} a
static solution with spherical symmetry was found. In many
respects it can be seen as a generalization of an extremal black
hole. Having infinitely many HS parameters all equal to each other
it generalizes the mass equals charge relation for usual extremal
black hole. At free level the $s=2$ Weyl tensor appears to be the
one of the Schwarzschild solution. The solution preserves some
amount of supersymmetries. In \cite{Bourdier:2014lya} it was
considered as embeddings in different susy HS models. In
\cite{Iazeolla:2011cb} the extremality condition was relaxed such
that HS parameters entering the solution become arbitrary thus
presenting a new family of black hole like solutions in HS theory
(see \cite{Iazeolla:2017dxc} for review).

While it seems not unlikely that the form of some of the obtained
exact solutions should be reconsidered in view of the locality
issue, some interesting information on their properties can still
be gained at free level. Particularly, a remarkable boundary
interpretation of the linearized version of HS extremal black hole
as a bi-local operator on the boundary was established in
\cite{David:2020fea}. The same authors have also observed that a
BPS-like pattern of the solution results in certain
UV-cancellations of two interacting such black hole states at
leading order.

A powerful method for generating HS analogs of black hole like
solutions at linear level rests on the observation from
\cite{Didenko:2009tc}. It was shown there that a vast class of
Petrov $D$-type solutions of General Relativity in four dimensions
that include Kerr, planar, {\it etc.} black holes are generated
with a single $AdS$ global symmetry parameter\footnote{In fact the
most general $D$ -- type Plebanski-Demianski metric can be
described this way by making the $AdS$ symmetry parameter not
hermitian conjugate \cite{DMVundone}.}. These naturally generalize
the double copy form to a multi copy one. From a twistor
standpoint this statement is equivalent to having some constant
rank-2 dual twistor that generates solutions to free massless
equations via the Penrose transform \cite{Didenko:2009td} (see
also \cite{Adamo:2017qyl}, \cite{White:2020sfn}). The resulting
Weyl tensors are of the generalized $D$ -- type. For the pure
gravity case a linearized Riemann tensor obtained this way is in
fact exact. The origin of this linearization is a hidden symmetry
of $D$ -- type metrics as they admit the so called Killing-Yano
tensor \cite{WP}, \cite{Floyd}. This property allows one naturally
incorporate the $s=2$ black hole solutions into the linearized HS
theory. The unfolded version of the Penrose transform well adopted
for HS analysis has been introduced in \cite{Gelfond:2008td},
\cite{Didenko:2009td}. It plays an important role in HS solution
generating techniques \cite{Didenko:2009td},
\cite{Iazeolla:2011cb}, \cite{Iazeolla:2017vng},
\cite{Aros:2017ror} as well as in $HS/CFT$ analysis
\cite{Didenko:2012tv}, \cite{Neiman:2017mel}.

The main results of our paper are the following. Using the proper
$AdS_4$ global symmetry parameter\footnote{See
\cite{Didenko:2015pjo} for incomplete classification of such
parameters.} (rank-2 dual twistor) we generate solutions for the
free bosonic HS fields $s\geq 0$ on the $AdS$ background via the
unfolded Penrose transform. These solve the so called 0-form
sector (Weyl module) of the linearized HS equations. We then
restore the HS potentials corresponding to the 1-form sector of HS
equations for the case of HS parity even model
($\eta=\bar\eta=1$). The obtained solutions are of $D$ -- type,
static and admit spatial planar symmetry similar to that of a
black brane. Moreover, the spin $s=2$ sector is exactly equal to a
black brane Weyl tensor in agreement with the well known result
from gravity.

While solving equations for the Weyl module $C$ using the
Penrose-like trick is not a problem, to recover the corresponding
HS potential sector $\go$ which is sourced by $C$ in a form
suitable for higher orders is not always an easy job. The problem
is somewhat equivalent to recover the metric from its Riemann
tensor. The procedure is gauge dependent. Even at free level the
result can be quite complicated compared to the form of the
original source $C$ (see e.g., \cite{Bolotin:1999fa},
\cite{Nagaraj:2019zmk}). For the planar solutions of our primary
interest we find surprisingly simple result for fields $\go$ in
the parity even HS model with $\eta=1$. On a technical side, one
reason why a simplification takes place is the existence of an
auxiliary $sp(2)$ flat connection induced by the generating
symmetry parameter (dual twistor) that results in a natural ansatz
for $\go$ which otherwise might be difficult to grasp. The use of
such a flat connection is beneficial at higher orders as well
\cite{DK} and plays a key role in our analysis.

Our solution is characterized by infinitely many parameters
attributed to different individual fields of given spin and
parameterized by an arbitrary analytic in the generating spinor
variables function. This function is polynomial in the case of a
finite amount of massless fields and non-polynomial otherwise,
which is a feature of the linearized approximation. An interesting
property of the obtained solution is the unique star-product Fock
projector that shows up as a factor for every spin $s$ field
within the Weyl module. Generally, Fock projectors play crucial
role in HS bulk-boundary analysis (see \cite{GY1},
\cite{Vasiliev:2012vf}, \cite{Didenko:2017lsn},
\cite{David:2020ptn}). Particularly, HS boundary-to-bulk
propagators are of this form \cite{Didenko:2012vh}. They as well
play a role of a probe data for possible non-localities within the
HS equations producing infinities for non-local
self-interaction.\footnote{Particularly, early analysis of HS
black brane at non-linear level has revealed some pathologies
within the non-local setup \cite{Dedundone}.} At the nonlinear
level Fock projectors tend to factorize (twisted)-adjoint HS
equations into left and right modules. It would be very
interesting to carry out higher order analysis of such solutions
especially within the context of bulk-boundary analysis of
\cite{Vasiliev:2012vf}, where a Fock projector naturally appears.
At quadratic level the HS corrections to the planar solution will
be analyzed in \cite{DK}.

The layout of the paper is as follows. In section \ref{Sec2} we
give a detailed description of algebraic properties of $AdS_4$
black brane. We show how its linearized nature results from hidden
symmetry attributed to a Killing-Yano tensor. It exists in the
vacuum background whether it is Minkowski or $AdS$ and along with
the Killing symmetry forms a global symmetry parameter that builds
up geometry of a black hole. We then give a condition on that
parameter to correspond to the planar symmetry. Then we briefly
review the Penrose transform that treats this parameter as a
rank-two twistor and allows one generating solutions for any spin
$s\geq 0$. We conclude this section with the description of the
brane induced $sp(2)$ flat connection that will play a
distinguished role in our analysis of the solutions of HS
equations. In section \ref{Sec3} the linearized Vasiliev HS
equations are reviewed and the unfolded version of the Penrose
transform is introduced. In section \ref{Sec4} we find solutions
of these equations and we conclude in section \ref{Sec5}. The
paper is supplemented with three Appendices that contain the
explicit form of the generating global symmetry parameter in the
Poincare chart, the derivation of the $sp(2)$ connection and,
finally, the details on derivation of our solution in the sector
of HS potentials.

\section{$AdS_4$ black brane}\label{Sec2}

The presence of negative cosmological constant affects drastically
classical topology theorems resulting in that black holes in
asymptotically $AdS$ space may have different horizon topologies.
Apart from the usual positive curvature horizon that Kerr solution
has there can be horizons of negative and zero curvatures. These
are the hyperbolic and planar black holes respectively. In fact,
in four dimensions an arbitrary genus Riemann surface horizon is
possible as one can quotient the hyperbolic horizon over a
discrete subgroup. We will be interested in a planar black hole
here the metric of which can be chosen in the following standard
form
\be\label{bbrane}
ds^2=-\ff{dr^2}{\Lambda r^2+M/r}+(\Lambda r^2+M/r)dt^2-\Lambda
r^2(dx^2+dy^2)\,,
\ee
where $\Lambda$ is the cosmological constant. When
\be
\Lambda<0
\ee
there is a horizon at
\be
r_0=\left(-\ff{M}{\Lambda}\right)^{\ff13}\,,
\ee
where $M>0$ is a massive parameter of the $AdS$ black brane.
Metric \eqref{bbrane} is manifestly time independent which implies
that the solution is stationary and in fact static. It as well has
no dependence on $x$ and $y$ meaning that there is another set of
isometries that leave the two dimensional spatial plane
\be
dl^2=dx^2+dy^2
\ee
invariant. These form the $iso(2)$ algebra and along with time
translation generate $u(1)\oplus iso(2)$ isometry algebra that can
be realized using vector fields
\begin{align}
&u(1): &&T=\ff{\p}{\p t}\,,\label{u1}\\
&iso(2): &&P_{1}=\ff{\p}{\p x}\,,\qquad P_{2}=\ff{\p}{\p
y}\,,\qquad L=x\ff{\p}{\p y}-y\ff{\p}{\p x}\label{iso}
\end{align}
with the commutation relations
\be\label{plan}
[T,P_{1,2}]=[T,L]=0\,,\qquad [P_{1},P_{2}]=0\,,\quad
[P_{1},L]=P_2\,,\quad [P_{2},L]=-P_1\,.
\ee
When $M=0$, \eqref{bbrane} reduces to the $AdS_4$ Poincare chart
\be\label{ads}
ds^2=\ff{1}{z^2}(-dt^2+dx^2+dy^2+dz^2)\,,\qquad
z=\ff{1}{\sqrt{-\Lambda} r}\,.
\ee
In these coordinates the spatial planar symmetry \eqref{iso} can
be realized as the centralizer of $T=\ff{\p}{\p t}\in so(3,2)$
which fact will be useful in what follows.

Four dimensional black holes in general and solution
\eqref{bbrane} in particular share the linearization property.
Namely, being exact solutions to non-linear Einstein equations
they at the same time satisfy the linearized ones. Moreover, the
non-linear corrections are satisfied as a consequence of the free
solutions. This property can be envisaged from the form of a black
hole Weyl tensor which is linear in $M$. The statement can be made
precise by means of the Kerr-Schild ansatz
\be\label{KS}
g_{mn}=g_{0\,mn}+\ff{M}{U}l_{m}l_{n}\,,
\ee
which decomposes black hole metric $g_{mn}$ in a sum of the $AdS$
background $g_{0\, mn}$ and the linear in $M$ fluctuation part
made of shear free geodesic congruence $l^{m}$
\be\label{KSV}
l_{m}l^{m}=0\,,\qquad l^{m}D_{m}l_{n}=0
\ee
and some scalar function $U$. Index contraction in \eqref{KSV} is
carried out with respect to either background or full metric. The
covariant derivative $D_m$ can be also attributed to either
metric. Equations \eqref{KSV} as well as the linearized Einstein
equations for \eqref{KS} guarantee that linear in $M$
approximation is exact. This behavior can be understood as
follows. Looking at black hole metric as some deformation of the
$AdS$ space, the linearization property suggests that certain
background geometry characteristics remain undeformed implying
that the black hole curvature is made of the $AdS$ background
remnants. This remnant appears to be a Killing-Yano tensor (see
\cite{Yasui:2011pr} for a comprehensive review) that $D$ -- type
metrics have which is present in $AdS$ (and flat) space-time and
remains unchanged upon deformation. Since the $AdS$ background
Killing-Yano does not depend on $M$ the black hole Riemann tensor
it provides has no any dependence on $M$ other than through a
linear in $M$ overall factor. The existence of the Killing-Yano
tensor in black hole geometry is often referred to as a hidden
symmetry, while its description is most accessible in the language
of two-component spinors.

\subsection{Hidden symmetry and linearization}

A convenient way to describe black holes that well captures their
algebraic properties is by using Cartan formalism. Consider Cartan
structure equations
\begin{align}
&\dr {\bf{w_{ab}}}+{\bf w}_{a}{}^{c}{\bf w}_{cb}=\bf{R}_{ab}\,,\label{Car1}\\
&D{\bf e}_{a}\equiv \dr {\bf e}_{a}+{\bf w}_{a}{}^{b}{\bf
e}_{b}=0\,.\label{Car2}
\end{align}
Here fields ${\bf w}_{ab}=-{\bf w}_{ba}$ and ${\bf e}_a$ are the
one-forms of Lorentz connection and vierbein respectively. The
two-form ${\bf R}_{ab}$ is the Riemann curvature. Indices are
contracted using the Minkowski metric $\eta_{ab}$.

The notion of hidden symmetry in a certain sense is a natural
generalization of Killing symmetries. Suppose one has a Lorentz
vector $t^{a}$. The action of covariant differential \eqref{Car2}
on it generally results in
\be
D t_{a}={\bf e}^{b}s_{ab}+{\bf e}^{b}n_{ab}\,,\label{Dt}
\ee
where $s_{ab}=s_{ba}$ and $n_{ab}=-n_{ba}$ are some
(anti)symmetric tensors that together form the most general right
hand side of \eqref{Dt}. If one of those tensors is absent then
the space-time may have a (hidden) symmetry. For example, if
$s_{ab}=0$, then $t^{a}$ is a Killing vector. Indeed, since
$n_{ab}$ is antisymmetric, equation \eqref{Dt} is equivalent to
\be\label{kill}
D_{a}t_{b}+D_{b}t_{a}=0\,,
\ee
which is just the Killing equation. If only the traceless part of
$s_{ab}$ is missing in \eqref{Dt}, then the resulting condition
would imply $t^a$ to be a conformal Killing vector. If instead
$n_{ab}=0$, then \eqref{Dt} is equivalent to $\p_{[a}t_{b]}=0$
which sets no restriction on geometry.

Similarly, one can consider more complicated tensor structures in
place of $t^a$, such as an antisymmetric tensor $t_{ab}=-t_{ba}$
for which one can write down
\be\label{Dt2}
Dt_{ab}={\bf e}^{c} r_{abc}+{\bf e}^{c} h_{ab,\,c}+{\bf
e}_{[a}n_{b]}\,,
\ee
where the right hand side of \eqref{Dt2} is decomposed in terms of
irreducible $so(3,1)$ traceless Young diagrams\footnote{In four
dimensions $r_{abc}$ can be dualized to a vector $v_a$ making the
first and third terms on the r.h.s. of \eqref{Dt2} equivalent. But
to set the nomenclature we are working in generic dimension for a
while.}
\begin{align}
&r_{abc}=r_{[abc]}\,,\\
&h_{ab,\, c}=h_{[ab],\, c}\,,\quad h_{[ab,\,c]}=0\,,\quad
\eta^{ac}h_{ab,\, c}=0\,.
\end{align}
$t_{ab}$ is said to be a rank-2 {Killing-Yano tensor (KY)} if
$h_{ab,\,c}=0$ and $n_{a}=0$ being a natural generalization for
Killing condition \eqref{kill}. Analogously, if only
$h_{ab,\,c}=0$ then such $t_{ab}$ is a {conformal KY}. If instead
$r_{abc}=0$ and $h_{ab,\,c}=0$, then $t_{ab}$ is called a {closed
conformal KY}. A higher rank totally antisymmetric tensor
corresponds to a higher-rank KY. There are no such structures in
four dimensions though.

The integrability requirement for \eqref{Car1}-\eqref{Car2}
$D^2\sim {\bf R_{ab}}$, $D{\bf R}_{ab}=0$ severely constrains any
symmetry, so that a generic space-time has no KY's. The simplest
example that admits KY symmetry is $AdS$ (or flat) space-time,
where it has a straightforward interpretation. To this end
consider system \eqref{Car1}-\eqref{Car2} for $AdS$ background
\begin{align}
&\dr {\bf{w_{ab}}}+{\bf w}_{a}{}^{c}{\bf w}_{cb}=\Lambda{\bf e}_{a}{\bf e_b}\,,\label{ads1}\\
&\dr {\bf e}_{a}+{\bf w}_{a}{}^{b}{\bf e}_{b}=0\,.\label{ads2}
\end{align}
A nice feature of this system is that it has local gauge symmetry
\begin{align}
&\gd{\bf w}_{ab}=D\gk_{ab}+\Lambda (v_a{\bf e}_{b}-v_b{\bf
e}_{a})\,,\label{gl1}\\
&\gd {\bf e}_{a}=Dv_{a}-\gk_{ab}{\bf e}^{b}\label{gl2}\,,
\end{align}
where $\gk_{ab}=-\gk_{ba}$ and $v_a$ are arbitrary space-time
dependent parameters. By setting $\gd{\bf w}_{ab}=0$ and $\gd {\bf
e}_{a}=0$ one fixes the $AdS$ global symmetry
\begin{align}
&Dv_{a}={\bf e}^{b}\gk_{ab}\,,\label{Ka}\\
&D\gk_{ab}=-\Lambda (v_{a}{\bf e}_b-v_{b}{\bf e}_a)\label{KYa}\,.
\end{align}
Here one identifies $v_a$ with a Killing vector, while $\gk_{ab}$
with a closed conformal KY. In $AdS$ the two fields go hand in
hand as parts of a global symmetry parameter, which can be
naturally written as an $so(3,2)$ covariantly constant matrix
$K_{IJ}=-K_{JI}$
\be\label{glob}
D_0 K_{IJ}=0\,,\qquad K_{IJ}=\begin{pmatrix} \gk_{ab} & \sqrt{-\Lambda} v_c \\
-\sqrt{-\Lambda} v_c & 0
\end{pmatrix}\,,
\ee
where
\be
D_0=d+{\bf w}_{IJ}\,,\qquad {\bf w}_{IJ}=\begin{pmatrix} {\bf w}_{ab} & \sqrt{-\Lambda} {\bf e}_c \\
-\sqrt{-\Lambda} {\bf e}_c & 0
\end{pmatrix}\,.
\ee
For metric \eqref{ads} with $x^{\mu}=(t,x,y,z)$ choosing vierbein
$e_{a,\,\mu}=\ff1z\eta_{a\mu}$ and taking Killing vector
\eqref{u1} $v^{\mu}=(1,0,0,0)$ and setting $\Lambda=-1$ for
convenience we find from \eqref{Ka}
\be\label{Kvec}
\gk_{ab}=\ff1z\left(
\begin{array}{cccc}
0 & 1 & 0 & 0\\
-1&0&0&0\\
0&0&0&0\\
0&0&0&0\\
\end{array}
\right)\,,\quad v_a=\ff1z(1,0,0,0)\,.
\ee
Black holes in $d=4$ and their $D$ -- type generalizations in
higher dimensions admit a nonzero closed conformal KY much as the
$AdS$ background  does. This means that \eqref{KYa} still applies
for black hole covariant derivative $D$, while \eqref{Ka} should
be modified because system \eqref{Ka}-\eqref{KYa} is consistent in
$AdS$ only. A consistent deformation of that system was studied in
\cite{Didenko:2009tc} using spinor language, where it was shown
that a general global symmetry parameter $K_{IJ}$ produces the
Carter-Plebanski family of $D$ -- type metrics which includes all
black hole solutions. Particularly, the Riemann tensor turns out
to be built out of closed conformal KY field $\gk_{ab}$ whereas
the massive parameter comes out as an overall factor. The fact
that \eqref{KYa} stays the same for black holes explains their
linearized nature, while smooth deformation of
\eqref{Ka}-\eqref{KYa} results in a certain integrating flow that
reconstructs black hole geometry in terms of the $AdS$ global
symmetry parameter \eqref{glob} \cite{Didenko:2009tc}. Different
parameters correspond to different types of black holes. The one
corresponding to the planar type has special algebraic properties
that we specify using spinors.

\subsubsection{Brane condition}
Isomorphism $so(3,2)\sim sp(4)$ allows us using the two-component
spinor language. In these terms symmetry parameter $K_{IJ}$
\eqref{glob} is equivalent to a symmetric $sp(4)$ matrix
$K_{AB}=K_{BA}$, $A,B=1,\dots ,4$. Let us also set cosmological
constant to a number for convenience, such that
\be\label{param} K_{AB}=\left(
\begin{array}{cc}
\gk_{\al\gb} & v_{\al\dgb}\\
v_{\gb\dal} & \bar{\gk}_{\dal\dgb}\\
\end{array}
\right)\,,\quad \gk_{\al\gb}=\gk_{\gb\al}\,,\quad
\bar{\gk}_{\dal\dgb}=\bar{\gk}_{\dgb\dal}\,,
\ee
whereas \eqref{Ka}-\eqref{KYa} reduce to
\begin{align}
&D\gk_{\al\gb}={\bf e}_{\al}{}^{\dgga}v_{\gb\dgga}+{\bf e}_{\gb}{}^{\dgga}v_{\al\dgga}\,,\label{param2a}\\
&D v_{\al\dal}={\bf e}_{\al}{}^{\dgga}\bar{\gk}_{\dal\dgga}+ {\bf
e}^{\gga}{}_{\dal}\gk_{\gga\al}\label{param2b}
\end{align}
or in a manifestly $AdS_4$ covariant way
\be
D_0 K_{AB}=0\,,
\ee
where $D_0=\dr+\Omega$,
\be\label{Om} \Omega_{AB}=\left(
\begin{array}{cc}
{\bf\go}_{\al\gb} & {\bf e}_{\al\dgb}\\
{\bf e}_{\gb\dal} & \bar{\bf\go}_{\dal\dgb}\\
\end{array}\right)\,.
\ee
Indices $\al,\dal=1,2$ are contracted with the help of $sp(2)$
forms $\gep_{\al\gb}=-\gep_{\gb\al}$ and
$\gep_{\dal\dgb}=-\gep_{\dgb\dal}$. Formally, system
\eqref{param2a}-\eqref{param2b} allows one generating the $D$ --
type (anti)self-dual parts of the Weyl tensor as follows
\be\label{Weyl}
C_{\al\gb\gga\gd}=\ff{M}{r^5}\gk_{(\al\gb}\gk_{\gga\gd)}\,,\qquad
\bar{C}_{\dal\dgb\dgga\dot\gd}=\ff{\bar M}{\bar
r^5}\bar\gk_{(\dal\dgb}\bar\gk_{\dgga\dot\gd)}\,,
\ee
where
\be\label{r}
r^2=-\ff12\gk_{\al\gb}\gk^{\al\gb}\,,\qquad \bar
r^2=-\ff12\bar\gk_{\dal\dgb}\bar\gk^{\dal\dgb}
\ee
and $M$ is an arbitrary parameter. Equations \eqref{Weyl}
correspond in particular to $AdS$ -- Kerr black hole and
encompasses generic $D$ -- case. A consistent deformation that
drives \eqref{param2a}-\eqref{param2b} away from $AdS$, $\hat
D^2\neq \Lambda{\bf e}{\bf e}$ looks as follows
\begin{align}
&\hat D\gk_{\al\gb}={\bf e}_{\al}{}^{\dgga}v_{\gb\dgga}+{\bf e}_{\gb}{}^{\dgga}v_{\al\dgga}\,,\label{defa}\\
&\hat D v_{\al\dal}=\bar\rho\, {\bf
e}_{\al}{}^{\dgga}\bar{\gk}_{\dal\dgga}+ \rho\, {\bf
e}^{\gga}{}_{\dal}\gk_{\gga\al}\,,\label{defb}
\end{align}
where $\rho$ and $\bar\rho$ are certain functions that depend on
deformation parameters (mass, NUT and electro-magnetic charges)
and on $r$ and $\bar r$ from \eqref{r} (see
\cite{Didenko:2009tc}). Note that \eqref{defa} remains unchanged
thus preserving $\gk_{\al\gb}$ as a KY symmetry and providing Weyl
tensors \eqref{Weyl} to be expressed in terms of background
fields.

While \eqref{Weyl} reproduces Weyl tensors for any $K_{AB}$ from
\eqref{param}, the case of a black brane corresponds (see
\cite{Didenko:2015pjo}) to real $M$ and
\be\label{brane}
K_{A}{}^{C}K_{CB}=0\,,\qquad
\det\gk=\ff12\gk_{\al\gb}\gk^{\al\gb}<0
\ee
with the following reality conditions imposed
\be
K_{AB}^{\dagger}=K_{AB}\,,\quad
\gk_{\al\gb}^{\dagger}=\bar\gk_{\dal\dgb}\,,\quad
v_{\al\dgb}^{\dagger}=v_{\gb\dal}\,.
\ee
In this case \eqref{Weyl} corresponds to a black brane which
metric is given by \eqref{bbrane}. Condition \eqref{brane} is
$sp(4)$ invariant with respect to the adjoint group action and
will be referred to as the brane condition. Let us also note the
$sp(4)$ subalgebra $\varepsilon$ that commutes with $K_{AB}$
\be
[\varepsilon, K]=0
\ee
spans planar symmetry \eqref{u1} and \eqref{iso}, where
$u(1)=\ff{\p}{\p t}$ part is generated by $K$ itself (see
Appendix). Component form of \eqref{brane} amounts to
\begin{align}
&\gk_{\al}{}^{\gga}\gk_{\gga\gb}+v_{\al}{}^{\dgga}v_{\gb\dgga}=0\,,\label{rel1}\\
&\gk_{\al}{}^{\gga}v_{\gga\dgb}=\bar{\gk}_{\dgb}{}^{\dgga}v_{\al\dgga}\,,\label{rel2}\\
&\bar\gk_{\dal}{}^{\dgga}\bar\gk_{\dgga\dgb}+v^{\gga}{}_{\dal}v_{\gga\dgb}=0\,,\label{rel3}
\end{align}
which entails
\be\label{br}
r^2=-\ff12
\gk_{\al\gb}\gk^{\al\gb}=-\ff12\bar{\gk}_{\dal\dgb}\bar{\gk}^{\dal\dgb}=\ff12
v_{\al\dal}v^{\al\dal}>0\,.
\ee
From \eqref{param2a} we also find that
\be\label{dr}
\dr r=\ff1r {\bf e}_{\al\dal}\gk^{\al\gga}v_{\gga}{}^{\dal}\,.
\ee
System \eqref{param2a}-\eqref{param2b} generates solutions to the
free $s=2$ equations. For example, the Kerr-Schild vector that
appears in \eqref{KS} as well as function $U$ can be expressed as
\be\label{KSk}
l_{\al\dal}=\ff{1}{r^2}\left(v_{\al\dal}+\ff1r\gk_{\al}{}^{\gb}v_{\gb\dal}\right)\,,\qquad
\ff1U=r\,.
\ee
It generates massless $s\geq 0$ multi copy solutions too
\cite{Didenko:2008va}. A comprehensive reason for this phenomenon
rests on the fact that \eqref{param2a}-\eqref{param2b} is the
rank-2 twistor equation. Therefore the Penrose transform can be
applied to produce a tower of massless solutions
\cite{Didenko:2009td}. While we are going to use the unfolded
version of the Penrose transform specified to $AdS$ in what
follows, let us now demonstrate briefly following
\cite{White:2020sfn} how the standard Penrose transform results in
$D$ -- type solutions in flat space-time.

\subsubsection{Flat space-time Penrose transform}
Suppose we are in flat space $\Lambda=0$ and therefore $D^2=0$ and
the vierbein can be chosen to be ${\bf e}^{\al\dgb}=\dr
x^{\al\dgb}$. A pair of spinors $Z^{A}=(\xi^{\al},\bar\xi_{\dal})$
is called a rank-1 twistor if the following condition is satisfied
\be\label{twistor}
D\xi^{\al}=-i{\bf e}^{\al\dal}\bar\xi_{\dal}
\ee
In the Cartesian reference frame it can be solved via
\be
\xi^{\al}=\xi^{\al}_{0}-i x^{\al\dal}\bar\xi_{0\,\dal}\,,\qquad
\bar\xi_{\dal}=\bar{\xi}_{0\,\dal}\,,
\ee
where $Z^{A}_{0}=(\xi^{\al}_0,\bar\xi_{0\,\dal})$ is $x$ --
independent. The incidence relation
\be\label{inc}
\xi^{\al}_{0}=i x^{\al\dal}_0\bar\xi_{0\,\dal}
\ee
then allows one establishing non-local correspondence between
points in space-time $x^{\al\dal}$ and points in twistor space
$Z^{A}$. Similarly, one can define a dual twistor
$Y_{A}=(\eta_{\al}, \bar\eta^{\dal})$ via hermitian conjugation of
\eqref{twistor}. Tensor product of relations \eqref{twistor} and
their conjugate results in an arbitrary rank twistor
$Z^{A_{1}\dots A_{n}}{}_{B_1\dots B_m}$. We can now define the
Penrose transform as a map from holomorphic twistor functions of
$Z^A=(u^{\al},\bar u_{\dal})$ into solutions of free massless
equations for (HS) Weyl tensors
\be\label{Pen1}
\bar{C}_{\dal_1\dots\,\dal_{2s}}=\oint_{\Gamma} \dr\bar
u^{\dgb}\bar u_{\dgb}\, \bar u_{\dal_1}\dots \bar u_{\dal_{2s}}
f(Z)\Big|_{CP^1}\,,
\ee
where the projection to $CP^1$ means that incidence relation
\eqref{inc} for twistor $Z^A$ is imposed. This implies that
function $f(Z)$ depends on $\bar u$ variable only
\be
f(Z)\Big|_{CP^1}=f(ix^{\al\dal}\bar u_{\dal}, \bar u_{\dgb})\,.
\ee
The contour $\Gamma$ is chosen to separate poles in such a way
that the integration makes sense. It is straightforward to check
now that the free massless spin $s$ equations hold
\be
\ff{\p}{\p x^{\al\dgb}}\bar
C^{\dgb}{}_{\dal_2\dots\,\dal_{2s}}=0\,.
\ee
To generate $D$ -- type free Weyl tensors it is sufficient to take
\be\label{pens}
f_{s+1}=\ff{1}{(K_{0\, AB}Z^{A}Z^{B})^{1+s}}\,,
\ee
where $K_{0\, AB}=K_{0\, BA}$ is some constant rank-2 dual
twistor. For a particular $K_{0}$ corresponding to Taub-NUT
case\footnote{In classification of \cite{Didenko:2015pjo} this
corresponds to $K_{A}{}^{C}K_{C}{}^{B}=\delta_{A}{}^{B}$.}
integration \eqref{Pen1} has been explicitly carried out in
\cite{White:2020sfn} with the final result being in agreement with
the general analysis of \cite{Didenko:2009tc}.

\subsection{Brane induced flat connection}

Getting back to $\Lambda<0$ case, most symmetric black holes of
spherical, planar or hyperbolic horizons are singled out by a
specific symmetry parameter $K$ that satisfies
\be\label{msym}
K_{A}{}^{C}K_{C}{}^{B}=\gep\,\delta_{A}{}^{B}\,,
\ee
where $ \gep=-1, 0, 1 $ correspondingly \cite{Didenko:2015pjo}.
Quite remarkably, in each case there exists an $sp(2)$ flat
connection that originates from mixing Lorentz components of
$AdS_4$ connection $\Omega$ \eqref{Om} and $K$ \eqref{param}. In
the planar case $\gep=0$ the form of such a connection is
especially simple\footnote{For the planar case the existence of
$sp(2)$ flat connection was independently confirmed by M.A.
Vasiliev.}. One can make sure using
\eqref{param2a}-\eqref{param2b} and \eqref{brane} (for more detail
see Appendix) that the following connection is $sp(2)$ flat
\be\label{wflat}
w_{\al\gb}=\go_{\al\gb}+\ff12({\bf
e}_{\al}{}^{\dgga}k_{\gb\dgga}+{\bf
e}_{\gb}{}^{\dgga}k_{\al\dgga})\,,
\ee
where
\be\label{k}
k_{\al\dal}=-\ff{1}{r^2}\gk_{\al}{}^{\gb}v_{\gb\dal}\,,\qquad
k_{\al}{}^{\dgga}k_{\gb\dgga}=\gep_{\al\gb}\,,\qquad
k^{\gga}{}_{\dal}k_{\gga\dgb}=\gep_{\dal\dgb}
\ee
and $\go_{\al\gb}$ is the holomorphic part of the $AdS$ Lorentz
connection. Similarly, one defines the dual connection
\be\label{wflatd}
\bar w_{\dal\dgb}=\bar\go_{\dal\dgb}+\ff12({\bf
e}^{\gga}{}_{\dal}k_{\gga\dgb}+{\bf
e}^{\gga}{}_{\dgb}k_{\gga\dal})\,.
\ee
This makes the following differential
\be
\nabla A_{\al\dal}=\dr
A_{\al\dal}-w_{\al}{}^{\gb}A_{\gb\dal}-\bar{w}_{\dal}{}^{\dgb}A_{\al\dgb}
\ee
indeed nilpotent
\be
\nabla^2=0\,,
\ee
implying the $sp(2)$ flatness condition
\be
\dr w_{\al\gb}-w_{\al}{}^{\gga}w_{\gb\gga}=0\,.
\ee
An advantage of this connection is that it makes the properly
rescaled components of \eqref{param} covariantly constant with
respect to $\nabla$
\be\label{const0}
\nabla\left(\ff{\gk_{\al\gb}}{r}\right)=0\,,\quad
\nabla\left(\ff{v_{\al\dgb}}{r}\right)=0\,,\quad
\nabla\left(\ff{\bar\gk_{\dal\dgb}}{r}\right)=0\,.
\ee
From \eqref{k} and \eqref{const0} it follows then
\be
\nabla k_{\al\dgb}=0\,.
\ee
Having two types of indices and being $\nabla$ -- constant we can
look at $k_{\al\dal}$ as a metric that converts dotted indices
into undotted ones and vise versa. Indeed, for any $A_{\dal}$ we
can define
\be\label{conv1}
A_{\al}:=k_{\al}{}^{\dgb}A_{\dgb}\,,
\ee
which entails from \eqref{k}
\be\label{conv2}
A_{\dal}=k^{\gga}{}_{\dal}A_{\gga}\,.
\ee
One should be cautious with signs though as this rule implies
\be\label{sign}
A^{\dal}B_{\dal}=-A^{\al}B_{\al}\,.
\ee
Components of \eqref{param} get unified under $k$ --
transformation in a sense that they transform into each other upon
index conversion
\be
k^{\gga}{}_{\dal}\gk_{\gga\al}=v_{\al\dal}\,,\quad
k^{\gga}{}_{\dal}v_{\gga\dgb}=\bar{\gk}_{\dal\dgb}\,,\quad
k_{\al}{}^{\dgga}v_{\gb\dgga}=\gk_{\al\gb}\,.
\ee
This suggests once one has metric $k_{\al\dal}$ the only
independent brane structure is, say, $\gk_{\al\gb}$, while the
rest result from it via index conversion. This fact is a mere
consequence of the  more general constraint \eqref{msym}.

Another observation is an analog of the vierbein postulate for
$\nabla$. One can check out the following identity
\be\label{frame}
\nabla {\bf E}_{\al,\,\gb}=0\,,
\ee
where we have introduced the one-form
\be
{\bf E}_{\al,\,\gb}:=\ff{1}{r}k_{\gb}{}^{\dgga}{\bf
e}_{\al\dgga}\,.
\ee
It will be convenient to decompose ${\bf E}_{\al,\,\gb}$ into its
symmetric and tracefull part
\be
{\bf E}_{\al,\,\gb}={\bf E}_{\al\gb}+\ff12\gep_{\al\gb}\, {\bf
E}\,,\qquad {\bf E}_{\al\gb}={\bf E}_{\gb\al}\,.
\ee
Both components are therefore covariantly constant
\be\label{const1}
\nabla{\bf E}_{\al\gb}=\nabla{\bf E}=0\,.
\ee
In terms of these fields the new connection reduces to
\be\label{post}
{w}_{\al\gb}=\go_{\al\gb}-r\, {\bf E}_{\al\gb}\,,\qquad \bar
w_{\dal\dgb}=\bar{\go}_{\dal\dgb}-r\, {\bf E}_{\dal\dgb}\,.
\ee
In view of \eqref{const0} $\ff1rK_{AB}$ is covariantly constant
with respect to connection $\nabla$. The only Lorentz scalar $r$
that system \eqref{param2a}-\eqref{param2b} has in this case is
not a constant as follows from \eqref{dr}
\be\label{d1r}
\dr\ff{1}{r}={\bf E}\,.
\ee
Properties \eqref{const0} as well as \eqref{post} play an
important role in solving HS equations within the black brane
ansatz especially greatly facilitating analysis at non-linear
level \cite{DK}.

\section{Higher-spin equations}\label{Sec3}

In this section we consider free bosonic HS equations using the
Vasiliev approach. The main reason for choosing this formalism is
that it makes higher-order analysis readily accessible on one hand
and  quite user friendly in four dimensions due to spinorial
language on the other. The price for that simplicity is an extra
set of fields that one has to deal with on top of the Fronsdal
ones. Speaking of exact solution this amounts to calculation of
auxiliary fields starting from free level.

HS equations of motion naturally group into the sector of gauge
fields and the sector of HS curvatures (Weyl tensors). The latter
admits an analog of the Penrose transform that gives a tool for
constructing solutions, while the former has a HS gauge freedom
that one should make use of properly. In practice, the analysis of
the gauge sector is most challenging.

In this approach fields are valued in higher-spin algebra, which
in $d=4$ is given by the Weyl algebra spanned by all polynomials
of the generating $Y_A=(y_{\al}, \bar y_{\dal})$ modulo the
relations
\be\label{com}
[y_\al, y_\gb]_*=2i\gep_{\al\gb}\,,\qquad [y_\al, \bar
y_{\dgb}]_*=0\,,\qquad [\bar y_{\dal}, \bar
y_{\dgb}]_*=2i\gep_{\dal\dgb}\,,
\ee
where star-product $*$ can be chosen to be the Moyal one
\be\label{exp}
f(Y)*g(Y)=f(Y)e^{i\gep^{AB}\overleftarrow{\p}_A\overrightarrow{\p}_B}
g(Y)\,.
\ee
In practice one uses the integral representation form for
exponential formula \eqref{exp}
\be\label{star}
f*g=\ff{1}{(2\pi)^4}\int\dr^2 u\dr^2\bar u\,\dr^2 v\dr^2\bar v
\,e^{iu_{\al}v^{\al}+i\bar u_{\dal}\bar v^{\dal}}f(y+u, \bar
y+\bar u)g(y+v, \bar y+\bar v)\,.
\ee
Vacuum of the theory satisfies the HS zero-curvature condition
\be\label{vaceq}
\dr W+W*W=0\,,
\ee
where $W=W(Y|x)$ is the one-form. The only polynomial solution of
that equation different from zero is the $AdS$ vacuum
\be\label{vac}
W_0=-\ff{i}{4}(\go^{\al\gb}y_{\al}y_{\gb}+\bar{\go}^{\dal\dgb}\bar{y}_{\dal}\bar{y}_{\dgb}+
2{\bf e}^{\al\dgb}y_{\al}\bar{y}_{\dgb})\,,
\ee
which upon feeding in \eqref{vaceq} results in Cartan structure
equations for $AdS_4$
\begin{align}\label{adsOmega}
&\dr\go_{\al\gb}-\go_{\al}{}^{\gga}\go_{\gga\gb}-{\bf e}_{\al}{}^{\dgga}{\bf e}_{\gb\dgga}=0\,,\\ \label{adsTetrad}
&\dr {\bf e}_{\al\dal}-\go_{\al}{}^{\gb}{\bf
e}_{\gb\dal}-\bar{\go}_{\dal}{}^{\dgb}{\bf e}_{\al\dgb}=0\,.
\end{align}
Any other $W$ satisfying \eqref{vaceq} inevitably contains fields
of any spin $s\geq 1$.

Gauge invariant sector describing HS Weyl tensors along with
matter fields ($s=0$ and $s=\ff12$) is encoded in the zero-form
$C(Y|x)$  that satisfies the so-called twisted-adjoint covariant
constancy condition
\be\label{twad}
\dr C+W_0*C-C*\pi(W_0)=0\,,
\ee
where $\pi(W_0)$ flips the sign of $y\bar y$ part in $W_0$
\eqref{vac} and more generally
\be\label{twist}
\pi f(y,\bar y)=f(-y,\bar y)\,.
\ee
Self-dual part of HS Weyl tensors is stored in $C(y,0)$, such that
the spin $s\geq 0$ component is
\be
C_{\al_1\dots\al_{2s}}=\ff{\p}{\p y^{\al_1}}\dots\ff{\p}{\p
y^{\al_{2s}}}C(y,\bar y)\Big|_{\bar y=0}\,.
\ee
Substituting \eqref{vac} into \eqref{twad} results in
\be\label{tw}
D C-i{\bf e}^{\al\dal}\left(y_{\al}\bar y_{\dal}-\ff{\p}{\p
y^{\al}}\ff{\p}{\p\bar y^{\dal}}\right)C=0\,,
\ee
where $D$ is the Lorentz covariant derivative
\be\label{lord}
D=\dr+\go^{\al\gb}y_{\al}\ff{\p}{\p
y^{\gb}}+\bar{\go}^{\dal\dgb}\bar{y}_{\dal}\ff{\p}{\p \bar
y^{\dgb}}\,.
\ee
Gauge fields are generated by one-form $w(y,\bar y|x)$ that
contains spin $s$ HS potential stored in $\go_{\al_1\dots
\al_{s-1},\,\dal_{1}\dots\dal_{s-1}}$ component of $w$ or,
equivalently, in the one singled out by
\be
y^{\al}\ff{\p}{\p y^{\al}} w=\bar y^{\dal}\ff{\p}{\p \bar
y^{\dal}}w=(s-1)w\,,
\ee
while other components are auxiliary. $w$ is sourced by the Weyl
module $C$. Its equation of motion can be obtained from the
Vasiliev system in the form\footnote{Note the (anti)holomorphic
dependence within $C$ on the right hand side of \eqref{OMS}. This
is the feature of ultra-locality that persists on any HS
background \eqref{vaceq} as well as at least at next interaction
order.} \cite{Vasiliev:1988sa}
\be\label{OMS}
D_0 w=\ff{i\eta}{4}{\bf e}^{\al\dgga}{\bf
e}^{\gb}{}_{\dgga}\ff{\p^2}{\p y^{\al}\p y^{\gb}}C(y,0|x)+
\ff{i\bar\eta}{4}{\bf e}^{\gga\dal}{\bf
e}_{\gga}{}^{\dgb}\ff{\p^2}{\p \bar y^{\dal}\p \bar
y^{\dgb}}C(0,\bar y|x)\,,
\ee
where
\be
D_{0}=\dr+[W_0, \bullet]_*=\dr+\go^{\al\gb}y_{\al}\ff{\p}{\p
y^{\gb}}+\bar{\go}^{\dal\dgb}\bar{y}_{\dal}\ff{\p}{\p \bar
y^{\dgb}}+ {\bf e}^{\al\dgb}\left(y_{\al}\ff{\p}{\p \bar
y^{\dgb}}+\bar{y}_{\dgb}\ff{\p}{\p y^{\al}}\right)\label{D0}
\ee
is the $AdS_4$ covariant derivative (c.f. \eqref{Om}). Parameter
$\eta$ is an arbitrary complex constant which unless $\eta=1$ or
$\eta=i$ breaks parity of the theory. Since $D_0^2=0$,
\eqref{vaceq} solutions to \eqref{OMS} are defined up to a gauge
freedom
\be\label{gfree}
w\sim w+D_0\gep\,.
\ee
A particular solution of \eqref{twad} $w$ and \eqref{OMS} $C$
breaks local gauge symmetry down to the leftover global one, which
is parameterized by $\gep_0(y,\bar y|x)$
\be\label{globsym}
D_0\gep_0=0\,,\qquad \gep_0*C-C*\pi(\gep_0)=0\,.
\ee

\subsection{Unfolded Penrose transform}
Let us now establish the $AdS$ HS analog of the flat Penrose
transform \eqref{Pen1} introduced in \cite{Gelfond:2008td},
\cite{Didenko:2009td}. Similar transformation has also been given
in \cite{Engquist:2005yt} and further elaborated in context of
relation between adjoint and twisted-adjoint HS modules in
\cite{Iazeolla:2008ix}. Its various aspects in context of solution
generating technique as well as in HS holography were detailed and
studied in recent years
\cite{Iazeolla:2011cb},\cite{Sundell:2016mxc},
\cite{Iazeolla:2017vng},  \cite{Iazeolla:2017dxc},
\cite{Aros:2017ror},\cite{Neiman:2017mel},
\cite{DeFilippi:2019jqq}, \cite{Aros:2019pgj},
\cite{Iazeolla:2020jee}.

To this end we define twistor\footnote{Having $sp(4)$ form
$\gep_{AB}$ we no longer distinguish between twistors and dual
twistors.} $T_A$ as an object that covariantizes \eqref{com} on
$AdS$ background
\be\label{twads}
D_0 T_A=0\,,\qquad [T_A, T_B]_*=2i\gep_{AB}\,,
\ee
where $D_0$ is given by \eqref{D0}. Equation \eqref{twads} can be
solved as
\be\label{T}
T_A=\Lambda_{A}{}^{B}Y_B\,,
\ee
where since $T_A$ preserves commutation relations, $\Lambda(x)$
should be an $Sp(4)$ group element equal to identity at some locus
point $x_0$, $\Lambda(x_0)=1$.

The Penrose transform is supposed to solve Weyl sector
\eqref{twad} of HS fields. In order to see this let us introduce
distributions $\gk_y=2\pi\delta^{2}(y)$ and $\bar\gk_{\bar
y}=2\pi\delta^2(\bar y)$ which properties mimic \eqref{twist}.
Using \eqref{star} one can show that
\be\label{del1}
F(y,\bar y)*\gk_{y}=\gk_{y}*F(-y,\bar y)\,,\qquad F(y,\bar
y)*\bar\gk_{\bar y}=\bar\gk_{\bar y}*F(y,-\bar y)\,,
\ee
where $F(y,\bar y)$ is an arbitrary function. Unlike standard
distributions that can not be squared, the product of
$\delta$-functions on non-abelian algebra \eqref{com} makes
perfect sense
\be\label{del2}
\gk_y*\gk_y=\bar\gk_{\bar y}*\bar\gk_{\bar y}=1\,.
\ee
To reveal the meaning of the introduced distributions one can find
using \eqref{star}
\be\label{Fourier}
F(y, \bar y)*\gk_y=\int\dr^2 u F(u, \bar y)e^{iu^{\al}y_{\al}}\,,
\ee
which shows that what $\gk_y$ does to a function is just the half
Fourier transform. This trick allows us solving \eqref{twad} in
terms of an arbitrary twistor function with constant coefficients
$F(T)$. Indeed, since from \eqref{twads} it follows that
\be
D_0 F(T)=\dr F+W_0*F-F*W_0=0\,,
\ee
and therefore to solve \eqref{twad} one needs an extra twist which
can be arranged using either $\gk_y$ or $\bar\gk_{\bar y}$. This
gives us, in particular,
\be\label{Pen}
C=F(T)*\gk_y
\ee
as a solution of the twisted-adjoint flatness condition
\eqref{twad}. Equation \eqref{Pen} and the similar conjugate one
with $\bar\gk_{\bar y}$ in place of $\gk_y$ will be referred to as
the unfolded Penrose transform. Explicitly, from \eqref{Fourier}
it follows
\be\label{expl}
C(y,\bar y|x)=\int\dr^2 u
F(\Lambda_{\al}{}^{\gb}u_{\gb}+\Lambda_{\al}{}^{\dgb}\bar
y_{\dgb},
\Lambda_{\dal}{}^{\gb}u_{\gb}+\Lambda_{\dal}{}^{\dgb}\bar
y_{\dgb})e^{iu^{\al}y_{\al}}\,.
\ee
The holomorphic part of spin $s$ Weyl tensor can be extracted by
setting $\bar y=0$ via
\be\label{comp}
C_{\al_1\dots\,\al_{2s}}\sim \int\dr^2\, u u_{\al_1}\dots
u_{\al_{2s}}F(\Lambda_{\al}{}^{\gb}u_{\gb},
\Lambda_{\dal}{}^{\gb}u_{\gb})\,.
\ee
Equation \eqref{comp} is an $AdS$ analog of the flat Penrose
transform \eqref{Pen1} and has much in common. There is however
some difference. The flat one is written down in the Cartesian
coordinates that allows one imposing explicit incidence relation
\eqref{inc} which enters the final result, while in $AdS$ we
stayed covariant implying that the $x$ -- dependence of twistor
$T$ is implicit and stored in \eqref{T}. Another difference is
while integration in \eqref{Pen1} is carried out along a closed
contour, in \eqref{comp} it goes across the two-dimensional plane.
Formally, one can rewrite \eqref{comp} in the form of a contour
integral by means of Stokes' formula
\be
\int \dr^2 u\,\p_{\al}g^{\al}(u)=\oint_{\Gamma} \dr u^{\al}u_{\al}
g(u)\,.
\ee
We prefer to keep integration over a plane for a reason. While the
contour presentation is convenient when dealing with functions
that have poles (see e.g. \cite{White:2020sfn}) which in addition
allows one cropping a particular spin out by adjusting the pole
degree, we stay with functions that are formally expandable in
$Y$'s in order to have control over star-product calculation. As a
result the unfolded Penrose transform \eqref{Pen} gives us a
solution for all spins at once rather than for a given one in
particular. Nevertheless,  we effectively find solutions for any
fixed spin doing this way, as we make advantage of in what
follows.

More importantly, as opposed to \eqref{Pen1}, its HS counterpart
\eqref{Pen} restores not only spin $s$ Weyl tensors but the whole
module $C$ that contains all on-shell derivatives of the former.
It allows us reaching global symmetries of the resulting solution
in a straightforward manner. Indeed, from \eqref{globsym},
\eqref{Pen} it follows that the leftover symmetries are those that
star commute with $F(T)$
\be\label{symcom}
[\gep, F(T)]_*=0\,.
\ee
For example, for solutions generated by some constant rank-2 field
$K_{0}^{AB}=K^{AB}(x_0)$ one picks
\be\label{exf}
F=F(K^{AB}(x_0)T_A T_B)=f(K^{AB}(x)Y_AY_B)\,,
\ee
where upon substitution \eqref{T}
\be
K^{AB}(x)=\Lambda_{C}{}^{A}K^{CD}(x_0)\,\Lambda_{D}{}^{B}
\ee
depends on $x$ via $Sp(4)$ group adjoint action that spreads
$K(x_0)$ across the $AdS$. Function \eqref{exf} generates HS Weyl
tensors via \eqref{comp} which have a certain space-time global
symmetry $\gep(Y|x)$ spanned by billinears in $Y$. To find this
symmetry one can take
$\gep=\xi^{AB}(x_0)T_{A}T_B=\xi^{AB}(x)Y_AY_B$, which manifestly
satisfies $D_0\gep=0$ provided $\xi^{AB}(x_0)$ are some constants.
Plugging it in \eqref{globsym} and using \eqref{symcom} results in
\be
[\xi^{AB}(x_0)T_AT_B, K^{CD}(x_0)T_CT_D]_*=0\,,
\ee
which using \eqref{twads} boils down to
\be\label{sym}
[\xi, K]_{AB}=0\,.
\ee
As expected, the global symmetry represented by parameter $\xi$
comes from the centralizer of $K$. In Particular, there are at
least two Killing vectors associated with $\xi_{1\,AB}=K_{AB}$ and
$\xi_{2\,AB}=K^{-1}_{AB}$ for generic $K$. This fact is in
agreement with the presence of the KY tensor in this case which
generates two Killing vectors \cite{Didenko:2009tc},
\cite{Yasui:2011pr}.

That global symmetries are easy to grasp within this approach is
thanks to the fact that the unfolded Penrose transform reproduces
the whole Weyl module $C(Y|x)$, where the symmetry action is
naturally realized via \eqref{globsym}. The standard Penrose
transform \eqref{Pen1} seemingly lacks direct access to global
symmetries of the solution it generates\footnote{We are grateful
to C. White for the discussion on this point.}. Let us also point
out that black hole- like fields generated with the aid of the
unfolded Penrose transform in \cite{Didenko:2009td} fall into a
double copy class of solutions on $AdS$. Their flat cousins were
considered from double copy perspective in e.g.
\cite{Monteiro:2014cda}.

\section{Solutions}\label{Sec4}

In this section we construct free static solutions with planar
symmetry for any spin $s\geq 0$. For $s=2$ the corresponding Weyl
tensor is given by \eqref{Weyl}, provided brane condition
\eqref{brane} is imposed. Our strategy is to first generate Weyl
module $C$ \eqref{twad} and then solve \eqref{OMS} that it
sources. Let us note that the Fronsdal fields result readily from
the Kerr-Schild ansatz. Indeed, as was shown in
\cite{Didenko:2008va} a spin $s\geq 0$ can be solved by
\be\label{fr}
\phi_{m_1\dots m_s}=\ff{m_s}{U}\,l_{m_1}\dots l_{m_s}\,,
\ee
where $m_s$ are arbitrary constants, while $l_m$ and $U$ are the
Kerr-Schild vector and the scalar function that $K_{AB}$ generates
via \eqref{KSk}. While \eqref{fr} looks neat and naturally
generalizes black hole metric \eqref{KS} still it is not most
suitable for higher order analysis since the auxiliary fields that
govern HS interaction have quite complicated form within the
Kerr-Schild setup. For parity even model with $\eta=1$ we find
different still simple form of the solution to \eqref{OMS}.

\subsection{Twisted-adjoint sector}

To find planar solutions for $C$ we use the unfolded Penrose
transform with twistor function \eqref{exf}. This has to be
supplemented with brane condition \eqref{brane} that guarantees
the required global symmetry \eqref{plan} in accordance with
\eqref{sym}. Let us take
\be
F=s \exp{\left(\ff{is}{2}K_{AB}Y^A Y^B\right)}
\ee
for \eqref{exf}, where $s$ is an arbitrary constant and $K_{AB}$
is the $AdS_4$ global symmetry parameter \eqref{param}. Simple
Gaussian integration in \eqref{Pen} and the use of
\eqref{rel1}-\eqref{rel3} gives the following result
\begin{align}\label{Fcal}
&F*\gk_{y}=s\int d^2 u'\, e^{\ff{is}{2}\varkappa_{\alpha\beta}
u'^\alpha u'^\beta}\exp{\left(\ff{i}{2s
r^2}\gk_{\al\gb}y^{\al}y^{\gb}-\ff{i}{r^2}\gk_{\al}{}^{\gb}v_{\gb\dal}y^{\al}\bar
y^{\dal}\right)}\sim\nn\\
&\sim\ff{1}{r} \exp{\left(\ff{i}{2s
r^2}\gk_{\al\gb}y^{\al}y^{\gb}-\ff{i}{r^2}\gk_{\al}{}^{\gb}v_{\gb\dal}y^{\al}\bar
y^{\dal}\right)}\,,
\end{align}
where $r$ is given in \eqref{br} and $\sim$ means the equality up
to an irrelevant at this stage complex phase factor attributed to
the fact that the gaussian integration in \eqref{Fcal} is carried
out with the complex quadratic form and therefore should be
completed via analytic continuation. Note the absence of
billinears in $\bar y$ in \eqref{Fcal} as they cancel out due to
brane condition \eqref{brane} being present otherwise. Similarly,
\be\label{barFcal}
F*\bar\gk_{\bar y}\sim\ff{1}{r} \exp{\left(-\ff{1}{2s
r^2}\bar\gk_{\dal\dgb}\bar y^{\dal}\bar
y^{\dgb}-\ff{i}{r^2}\gk_{\al}{}^{\gb}v_{\gb\dal}y^{\al}\bar
y^{\dal}\right)}\,.
\ee
Now, since $s$ is arbitrary one can integrate \eqref{Fcal} and
\eqref{barFcal} over $s$ with any measure\footnote{For example, a
spin $m$ field can be extracted via residue $\oint\dr s s^{2m-1}
F*\gk_y$ around zero. Note that the contour integration can not be
interchanged with the star product one in this case.} $\rho(s)$
still producing solutions for $C$ which can be written down in the
following form
\be\label{Csol}
C(y,\bar
y|x)=\ff{1}{r}f\left(\ff{1}{2r^2}\gk_{\al\gb}y^{\al}y^{\gb}\right)e^{-\ff{i}{r^2}\gk_{\al}{}^{\gb}v_{\gb\dal}y^{\al}\bar
y^{\dal}}+\ff{1}{r}\bar
f\left(\ff{1}{2r^2}\bar\gk_{\dal\dgb}\bar{y}^{\dal}\bar{y}^{\dgb}\right)e^{-\ff{i}{r^2}\gk_{\al}{}^{\gb}v_{\gb\dal}y^{\al}\bar
y^{\dal}}\,,
\ee
where $f$ is an arbitrary (complex) analytic in $y$ function.
Since $r$ is real \eqref{br}, solution \eqref{Csol} clearly
satisfies usual HS reality condition (see eq. \eqref{twist} for
the definition of $\pi$)
\be
C^{\dagger}=\pi(C)\,.
\ee
Let us stress that the procedure of obtaining \eqref{Csol} via
Penrose transform \eqref{Pen} should be viewed as a guiding
principle at best. We are not going to use function $F(Y)$
anywhere in what follows as it can well happen to be nonanalytic
in $Y$'s (see for example the analogous \eqref{pens}). Indeed, it
is $f(y)$ rather than $F(Y)$ should be analytic in order to
correspond to HS fields. Moreover, $f(y)$ is polynomial for finite
amount of spins. In this case however the corresponding $F(Y)$ is
not analytic which makes it problematic to work with within the HS
algebra\footnote{This is a general phenomenon immanent to the
unfolded Penrose transform for it relates an infinite dimensional
twisted-adjoint HS module to a finite-dimensional adjoint one. In
practice it transforms polynomials into distributions and vice
versa. An infinite tower of spins is necessary to have regular
functions on both sides of the Penrose transform.}. The fact that
\eqref{Csol} satisfies \eqref{tw} for any $f(y)$ can be checked
directly without any reference to the Penrose transform.

The Weyl tensors \eqref{Csol} generates are of $D$ -- type
\begin{align}
&C_{\al_1\dots
\,\al_{2s}}=\ff{m_s}{r^{2s+1}}\gk_{(\al_1\al_2}\dots\gk_{\al_{2s-1}\al_{2s})}\,,\\
&\bar C_{\dal_1\dots \,\dal_{2s}}=\ff{\bar
m_s}{r^{2s+1}}\bar\gk_{(\dal_1\dal_2}\dots\bar\gk_{\dal_{2s-1}\dal_{2s})}\,,
\end{align}
where the mass-like parameters $m_s$ are
\be
\quad m_s=\left(\ff{\p}{\p x}\right)^s f\left(\ff
x2\right)\Big|_{x=0}\,.
\ee
Particularly, the scalar $s=0$ corresponds to $\Delta=1$ boundary
behavior
\be
C_{\Delta=1}=\ff{m_0}{r}\,.
\ee

While $m_{s>0}$ can be complex corresponding to both the magnetic-
and electric- like HS parameters, we restrict ourselves to the
case of real $m_s=\bar m_s$ which implies $f=\bar f$. For $s=2$
the respective Weyl tensors are of $AdS_4$ black brane
\eqref{bbrane} of mass $M=m_2$. Note, that $f$ which reproduces a
single given spin $s$ is just $f(x)=m x^s$. Being no more than
polynomial, all information on the Weyl tensor on-shell
derivatives is stored in the exponential of \eqref{Csol}, which
turns out to be the same for any spin thanks to the degenerate
brane condition \eqref{brane}. This exponential has a remarkable
projector property which becomes crucial at non-linear level
\cite{DK}.

Before going into the details let us first make use of the
convention introduced in \eqref{conv1}-\eqref{conv2}. We note that
what appears in the exponential of \eqref{Csol} is exactly
\eqref{k}. This leads us to define
\be
\bar y_{\al}:=k_{\al}{}^{\dgb}\bar{y}_{\dgb}\,.
\ee
Having this notation, solution \eqref{Csol} takes the form
\be\label{Csol1}
C=\ff{1}{r}(f(x)+f(\bar x))e^{i y_{\al}\bar y^{\al}}\,,\quad
x=\ff{1}{2r^2}\gk_{\al\gb}y^{\al}y^{\gb}\,,\quad \bar
x=\ff{1}{2r^2}\gk_{\al\gb}\bar y^{\al}\bar y^{\gb}\,.
\ee
Now, one recognizes the Fock projector
\be
e^{i y_{\al}\bar y^{\al}}*e^{i y_{\al}\bar y^{\al}}=e^{i
y_{\al}\bar y^{\al}}\,,\qquad e^{i y_{\al}\bar y^{\al}}*\pi (e^{i
y_{\al}\bar y^{\al}})=\gd^{(2)}(0)\,,
\ee
which forms a Fock vacuum state for the following creation and
annihilation operators
\be
a^{\pm}=y\pm\bar y\,,\qquad a^-*e^{i y_{\al}\bar y^{\al}}=e^{i
y_{\al}\bar y^{\al}}*a^+=0\,.
\ee

In this respect let us note that projectors greatly facilitate
non-linear HS analysis and for this reason are often introduced
for solving the non-linear Vasiliev equations. In
\cite{Didenko:2009td}, where spherically symmetric solution was
considered, the mass-like HS parameters for infinitely many HS
fields were fine tuned to form a projector. This is in sharp
contrast with our planar case, where the Fock projector shows up
for each spin $s$ field right away.

Bearing in mind that Fock projector tends to self-reproduce at
nonlinear level (see \cite{Didenko:2017lsn} where it was noted at
quadratic level in the $AdS/CFT$ context) we rewrite \eqref{tw} so
as to separate the projector dependence. To this end we decompose
Lorentz covariant derivative \eqref{lord} by separating part that
contains $sp(2)$ flat connection \eqref{post} the oscillator
representation of which reads
\be
\nabla=D+{\bf w}^{\al\gb}y_{\al}\ff{\p}{\p y^\gb}+\bar {\bf
w}^{\dal\dgb}\bar y_{\dal}\ff{\p}{\p y^{\dgb}}=D-r {\bf
E}^{\al\gb}\left(y_{\al}\ff{\p}{\p y^{\gb}}-\bar
y_{\al}\ff{\p}{\p\bar y^{\gb}}\right)\,,
\ee
where we used
\be
\ff{\p}{\p\bar y^{\dal}}=-k^{\gb}{}_{\dal}\ff{\p}{\p\bar
y^{\gb}}\,.
\ee
This way twisted-adjoint equation \eqref{tw} reduces to
\be\label{tw1}
\nabla C+ir {\bf E}^{\al\gb}(y+i\bar \p)_{\al}(\bar
y-i\p)_{\gb}C+\ff{ir}{2}{\bf E}(y_{\al}\bar
y^{\al}+\p_{\al}\bar\p^{\al})C=0\,.
\ee
We can now define a current module\footnote{A very similar
decomposition was used in \cite{Vasiliev:2012vf} for HS
bulk-boundary analysis, where the corresponding $T$ appears to
describe $3d$ boundary currents. We expect the analogy to be far
reaching and hence borrow the terminology.} $T(y,\bar y)$ by
separating the Fock projector
\be
C(y,\bar y)=T(y,\bar y)e^{iy_{\al}\bar y^{\al}}\,.
\ee
Using that $\nabla e^{iy_{\al}\bar y^{\al}}=0$ we find from
\eqref{tw1}
\be\label{Teq}
\nabla T+ir {\bf E}^{\al\gb}\bar \p_{\al}\p_{\gb}T-\ff{r}{2}{\bf
E}(y^{\al}\p_{\al}+\bar y^{\al}\bar
\p_{\al}-i\p_{\al}\bar\p^{\al}+2)T=0\,.
\ee
It is straightforward to check that $T=\ff1r (f(x)+f(\bar x))$
solves \eqref{Teq} for any function $f$. There is however another
solution with planar symmetry which is not on the list
\eqref{Csol1}. Indeed,
\be
T_{\Delta=2}=\ff{1}{r^2}(1+i y_{\al}\bar y^{\al})
\ee
corresponding to a scalar field with the alternative boundary
condition $\Delta=2$ does satisfy \eqref{tw1}. It was not captured
by the unfolded Penrose transform \eqref{Fcal} and we can add it
to \eqref{Csol1} if necessary for the nonlinear analysis
\be
C_{\Delta=2}=\ff{m_0'}{r^2}\,.
\ee

\subsection{Adjoint sector}
Solutions to \eqref{OMS} is much harder to find. The unfolded
Penrose transform does not act in this sector, while equation
\eqref{OMS} is not homogeneous being sourced by primary components
of the Weyl module $C$. Therefore, the solution procedure is a
reconstruction of HS potentials in terms of curvatures. The result
is gauge dependent as is seen from \eqref{gfree} and can be quite
complicated even for plane wave- like solutions,
\cite{Bolotin:1999fa}, \cite{Nagaraj:2019zmk}. In addition, what
one needs is not only Fronsdal components of $w(y,\bar y)$, which
form is quite simple in the Kerr-Schild representation \eqref{fr},
but the whole adjoint module that contains auxiliary fields
related to the Fronsdal ones via derivatives. These auxiliary
fields contribute to interactions within the unfolded approach. In
principle one can start from the physical components given by
\eqref{fr} and restore  all auxiliary fields one by one. While
doable in principle the result turns out to be not particularly
encouraging. Instead, inspired by flat connection \eqref{post} we
provide an ansatz that works in HS A-model corresponding to
$\eta=\bar\eta=1$ in \eqref{OMS} and brings us to the solution for
the whole module $w$ in a simple form.

Our strategy is as follows. Since the only variable which is not
constant with respect to connection $\nabla$ is $r$ \eqref{d1r} we
can rewrite \eqref{OMS} in terms of $\nabla$ and then propose an
ansatz for $w(y,\bar y|x)$ that depends on $\nabla$- covariantly
constant fields \eqref{const0}, \eqref{const1} as well as on $r$
thus reducing the space-time partial differential equation to the
ordinary one with respect to $r$.

Proceeding this way we first rewrite \eqref{OMS} in $\nabla$-
covariant way by expressing \eqref{D0} as
\be
D_0=\nabla+r{\bf E}^{\al\gb}(y-\bar y)_{\al}(\p+\bar\p)_{\gb}-\ff
r2{\bf E}(y^{\al}\bar\p_{\al}+\bar y^{\al}\p_{\al})\,.
\ee
This leads us to the following form\footnote{By introducing
$a^{\pm}=y\pm\bar y$, equation \eqref{omsn} takes a form very
similar to the one that arises in bulk-boundary analysis of
\cite{Vasiliev:2012vf}.} of \eqref{OMS}
\begin{align}\label{omsn}
&\left(\nabla+r{\bf E}^{\al\gb}(y-\bar
y)_{\al}(\p+\bar\p)_{\gb}-\ff
r2{\bf E}(y^{\al}\bar\p_{\al}+\bar y^{\al}\p_{\al})\right)w=\\
=&\ff{i r^2}{4}{\bf E}^{\al\gga}{\bf
E}_{\gga}{}^{\gb}\left(\bar\eta\bar\p_{\al}\bar\p_{\gb}C(0,\bar
y)-\eta\p_{\al}\p_{\gb}C(y,0)\right)+\ff{ir^2}{4}{\bf
E}^{\al\gb}{\bf E}\left(\eta\p_{\al}\p_{\gb}C(y,0)+
\bar\eta\bar\p_{\al}\bar\p_{\gb}C(0,\bar y)\right)\,.\nn
\end{align}
Substituting \eqref{Csol1} into \eqref{omsn} one arrives at
\begin{align}\label{weq}
&\left(\nabla+r{\bf E}^{\al\gb}(y-\bar
y)_{\al}(\p+\bar\p)_{\gb}-\ff
r2{\bf E}(y^{\al}\bar\p_{\al}+\bar y^{\al}\p_{\al})\right)w=\\
=&\ff{i r}{4}{\bf E}^{\al\gga}{\bf
E}_{\gga}{}^{\gb}\left(\bar\eta\bar\p_{\al}\bar\p_{\gb} f(\bar y)
-\eta\p_{\al}\p_{\gb}f(y)\right)+\ff{ir}{4}{\bf E}^{\al\gb}{\bf
E}\left(\eta\p_{\al}\p_{\gb}f(y)+ \bar\eta\bar\p_{\al}\bar\p_{\gb}
f(\bar y)\right)\,.\nn
\end{align}
To solve this equation we propose the following ansatz
\be\label{anz}
w={\bf E}^{\al\gb}\int_{0}^{1}\dr\tau
\rho(\tau)\p_{\al}\p_{\gb}g(\tau y+(1-\tau)\bar y)\,,
\ee
where $\p$ differentiates the argument of $g$,
\be
\p_\al g(\xi):=\ff{\p}{\p \xi^{\al}}g(\xi)
\ee
and $\rho(\tau)$ is an unspecified function. The idea behind
\eqref{anz} is that upon substitution into \eqref{weq} the result
may acquire a form of a total derivative with respect to $\tau$,
such that at $\tau=1$ it leads to $\p_{\al}\p_{\gb}f(y)$, and to
$\bar\p_{\al}\bar\p_{\gb}f(\bar y)$ at $\tau=0$ correspondingly,
thus matching the right hand side. The particular dependence on
frame fields ${\bf E}_{\al\gb}$ and ${\bf E}$ is also a part of
our ansatz motivated by the presence of second derivatives on the
right hand side of \eqref{weq}.

Let us check if \eqref{anz} works. When acting on $w$, the left
hand side of \eqref{weq} splits  into two two-form sectors ${\bf
E}^{\al\gga}{\bf E}_{\gga}{}^{\gb}$ and ${\bf E}^{\al\gb} {\bf
E}$. The first one that has contribution from the second term on
the l.h.s. of \eqref{weq} only reads
\be\label{eq1}
{\bf E}^{\al\gb}(y-\bar y)_{\al}\left(\ff{\p}{\p y}
+\ff{\p}{\p\bar{y}}\right)_{\gb} {\bf
E}^{\gga\gd}\int_{0}^{1}\dr\tau \rho(\tau)\p_{\gga}\p_{\gd}g={\bf
E}^{\al\gb}{\bf E}^{\gga\gd}(y-\bar y)_{\al} \int_{0}^{1}\dr\tau
\rho(\tau)\p_{\gb}\p_{\gga}\p_{\gd}g\,.
\ee
With the aid of the Schouten identities
\be
{\bf E}^{\al\gb}{\bf E}^{\gga\gd}=\ff14(\gep^{\al\gga}{\bf
H}^{\gb\gd}+\gep^{\gb\gga}{\bf H}^{\al\gd}+\gep^{\al\gd}{\bf
H}^{\gb\gga}+\gep^{\gb\gd}{\bf H}^{\al\gga})\,,\qquad {\bf
H}^{\al\gb}=-{\bf E}^{\al\gga}{\bf E}_{\gga}{}^{\gb}
\ee
\eqref{eq1} is further reduced to
\be
-\ff12 {\bf H}^{\gb\gga}(y-\bar
y)^{\al}\p_{\al}\int_{0}^{1}\dr\tau
\rho(\tau)\p_{\gb}\p_{\gga}g=-\ff12 {\bf
H}^{\al\gb}\int_{0}^{1}\dr\tau
\rho(\tau)\ff{\p}{\p\tau}\p_{\al}\p_{\gb}g\,,
\ee
where in the last line we made use of the following identity
\be
(y-\bar y)^{\al}\p_{\al}g(\tau y+(1-\tau)\bar y)=\ff{\p}{\p\tau}
g(\tau y+(1-\tau)\bar y)\,.
\ee
We see that the result indeed acquires a form of a total
derivative. Partial integration results in
\be
\int_{0}^{1}\dr\tau
\rho(\tau)\ff{\p}{\p\tau}\p_{\gb}\p_{\gb}g=\rho(1)\p_{\gb}\p_{\gb}g(y)-\rho(0)\bar\p_{\gb}\bar
\p_{\gb}g(\bar y)-\int_{0}^{1}\dr\tau
\rho'(\tau)\p_{\gb}\p_{\gb}g\,.
\ee
In order to match the right hand side of \eqref{weq} one sets
$g=f$, $\rho=-1$ and $\eta=\bar\eta=1$ thus having
\be\label{wsol}
w=-\ff{i}{2}{\bf
E}^{\al\gb}\int_{0}^{1}\dr\tau\p_{\al}\p_{\gb}f(\tau
y+(1-\tau)\bar y)\,.
\ee
Sector ${\bf E}^{\al\gga}{\bf E}_{\gga}{}^{\gb}$ is perfectly
satisfied by our ansatz, provided $\eta=\bar\eta$ which
corresponds to the HS $A$-model case. On the other hand we have no
freedom left within \eqref{anz}, which suggests that the remaining
sector ${\bf E}^{\al\gb}{\bf E}$ is either satisfied by default or
the ansatz does not go through.

To check out if \eqref{wsol} solves ${\bf E}^{\al\gb}{\bf E}$
sector is less trivial. In fact it does not for generic $f(\xi)$,
but in the case of
\be\label{f}
f(\xi)=f\left(\ff{1}{2r^2}\gk_{\al\gb}\xi^{\al}\xi^{\gb}\right)\,,
\ee
which precisely corresponds to Weyl module \eqref{Csol1}, it does
go through. Being somewhat technical we leave this verification
for Appendix, still noting here that a reason other than just a
mere coincidence for why \eqref{wsol} satisfies another constraint
from sector ${\bf E}^{\al\gb}{\bf E}$ is not entirely clear to us.

As a result, the adjoint module is solved by simple formula
\eqref{wsol}. It corresponds to the gauge where $y$ and $\bar y$
enter in a totally symmetric way. Indeed, from \eqref{wsol} and
\eqref{f} one can obtain the following component form for $w$
\be\label{totsym}
w_{\al(m), \gb(n)}=\ff{1}{r^{2+m+n}}T_{\al(m)\gb(n)\gga\gd}{\bf
E}^{\gga\gd}\,,
\ee
where $T_{\al_1\dots\,\al_{2k}}\sim
\gk_{(\al_1\al_2}\dots\gk_{\al_{2k-1}\al_{2k})}$ is totally
symmetric. Hence, of gauge that results in such $w$ we can refer
to as of the symmetric one.

Summarizing here our findings, the HS potentials that corresponds
to Weyl module \eqref{Csol1} have the following form
\be
w=-\ff{i}{2}{\bf
E}^{\al\gb}\int_{0}^{1}\dr\tau\p_{\al}\p_{\gb}f\left(\ff{1}{2r^2}\gk_{\gga\gd}
(\tau y+(1-\tau)\bar y)^{\gga}(\tau y+(1-\tau)\bar
y)^{\gd}\right)\,,
\ee
where we recall that here $\p$ differentiates with respect to
$\tau y+(1-\tau)\bar y$. To get back to original variables
\eqref{param} one substitutes
\be
\bar y_{\al}=-\ff{1}{r^2}\gk_{\al}{}^{\gb}v_{\gb}{}^{\dal}\bar
y_{\dal}\,,\qquad {\bf
E}_{\al\gb}=-\ff{1}{2r^3}(\gk_{\al}{}^{\gga}v_{\gga}{}^{\dgga}{\bf
e}_{\gb\dgga}+\gk_{\gb}{}^{\gga}v_{\gga}{}^{\dgga}{\bf
e}_{\al\dgga})\,.
\ee
Let us stress that the form of the final result for $w$ is hard to
anticipate without being guided by the auxiliary $sp(2)$ flat
connection \eqref{wflat}. It plays an important role in our
derivation of $w$ and remains so at the interaction level
considered in \cite{DK}.

\section{Conclusion}\label{Sec5}
We have initiated a search for static solutions in $d=4$
higher-spin gauge theory that have planar symmetry. Owing to the
standard gravity examples we expect such solutions may correspond
to a thermal state of a boundary theory. The pursuit was
encouraged by the solution generating technique of
\cite{Didenko:2009tc}, \cite{Didenko:2009td} that produces black
hole- like solutions from the $AdS$ global symmetry parameter
$K_{AB}$ for all spins at once. From the twistors standpoint this
parameter is a rank-2 twistor and therefore can be fed into the
Penrose transform yielding the Petrov $D$ -- type solutions. Since
HS equations are based on the Vasiliev unfolded approach the
Penrose transform gets modified accordingly and we call such a
modification introduced in \cite{Gelfond:2008td},
\cite{Didenko:2009td} the unfolded Penrose transform. We provide
details on that modification and compare it with the standard one.
A major difference is the unfolded transform not only reproduces
solutions to free HS Weyl tensors but as well to their all
descendants which are on-shell derivatives thereof. This gives us,
in particular, the access to the leftover global symmetries of the
solutions it generates. In the planar case of our interest this
implies the generating parameter should be degenerate satisfying
\eqref{brane}.

Property \eqref{msym} gives rise a certain $sp(2)$ flat connection
that makes the properly rescaled Lorentz components of parameter
$K_{AB}$ covariantly constant tensors with respect to this
connection. That made us use the $sp(2)$ covariant formalism in
our analysis. Using the unfolded Penrose transform we have found a
class of solutions of the HS Weyl module parameterized by a single
arbitrary analytic in $y$'s function that encodes an infinite set
of HS mass-like parameters. These include a free scalar with
$\Delta=1$ boundary condition and $s\geq 1$ gauge fields that have
the double copy origin. The function is of homogeneity degree $s$
for a single spin $s$ field. While the solution for $s=0$ with
$\Delta=2$ boundary condition was not captured by our Penrose
transform, it nevertheless naturally exists. For $s=2$ our
solution corresponds to the $AdS_4$ black brane.

Remarkable property of the obtained solution is that it contains
one and the same Fock-type projector for any spin $s\geq 0$.
Projectors are known to appear in the $AdS/CFT$ context as
bulk-to-boundary propagators \cite{GY1}, \cite{Didenko:2012vh},
\cite{Didenko:2012tv}. It is conceivable that at the full
non-linear level it may play a role of some thermal state. The
nonlinear analysis gets especially interesting in the presence of
projectors as they tend to survive in interactions and eventually
decouple from HS equations \cite{Didenko:2017lsn}. This is
especially relevant in context of (non)locality of HS
interactions. Indeed, the decoupling of the projector leaves one
with HS equations that contain no more than polynomial
interactions for the states bounded in spin.

While the obtained solutions are by construction of $D$ Petrov
type, it is also interesting to check whether they stay so in
interaction. It would be also interesting to see whether the pure
$s=2$ field corresponding to $AdS$ black brane at leading order
induces higher spins in interaction. In order to be able to answer
these kind of questions one has to have complete solution at free
level that includes both the sector of gauge invariant curvatures
and the one of gauge fields. While the former comes about from the
Penrose transform, the latter can be difficult to find. One of the
main result of this paper is a simple formula for HS gauge fields
\eqref{wsol} that we were able to derive in parity even HS theory
$\eta=\bar \eta=1$. It makes a further quadratic analysis readily
accessible \cite{DK}. It would be also interesting to know if
there is as simple generalization of formula \eqref{wsol}
available for parity broken theory.

A great deal of simplification in our derivation of HS gauge
fields results from the emergent $sp(2)$ flat connection. On a
practical side this implied that the space-time partial derivative
equations of motion get reduced to the ordinary one along the
radial variable $r$. Our result for the field components of HS
connections $w_{\al(m),\,\dgb(n)}$ from $w(y,\bar y)$ turns out to
be totally symmetric with respect to permutations in both group of
indices. Finally, let us note that the flat connection trick can
be also useful in different context. In particular, the
bulk-to-boundary propagator for gauge fields written in
\cite{Didenko:2017lsn}, \cite{Sezgin:2017jgm} can be found using
similar trick.

\section*{Acknowledgments}
We are grateful to Chris White for useful discussions and to Mitya
Ponomarev and M.A. Vasiliev for valuable comments on the draft of
the paper. VD acknowledges participation of Oleg Shaynkman at the
early stage of this work. Finally, we would like to thank the
anonymous Referee for many valuable comments. The research was
supported by the RFBR grant No 20-02-00208.

\section*{Appendix A. Explicit form of the generating parameter}
The $AdS$ global symmetry parameter that generates planar
solutions have particular convenient form in Poincare coordinates
\eqref{ads}. Its $so(3,1)$ tensor components originating from
Killing vector $\ff{\p}{\p t}$ are given in \eqref{Kvec}. The
$sp(4, R)$ spinor realization is easy to obtain by introducing the
$so(3,2)$ $\gga$-matrices $\{\gga_{I}, \gga_{J}\}=2\eta_{IJ}$,
which can be taken to be the following $\gga_I=(\gga_a, \gga_5)$
\be
\gamma_0=\begin{pmatrix} -i \sigma^2 & 0 \\ 0 & i \sigma^2
\end{pmatrix}, \ \gamma_1=\begin{pmatrix} \sigma^1 & 0 \\ 0 &
\sigma^1 \end{pmatrix}, \ \gamma_2=\begin{pmatrix} \sigma^3 & 0 \\
0 & \sigma^3 \end{pmatrix}, \ \gamma_3=\begin{pmatrix} 0 & i
\sigma^2  \\ -i \sigma^2 & 0 \end{pmatrix},
\ee
\be
\gga_5=\gga_0\gga_1\gga_2\gga_3=\begin{pmatrix} 0 & i\gs^2 \\
i\gs^2 & 0 \end{pmatrix}\,.
\ee
We can now define
\be
K_{AB}=\ff12K_{IJ}(\gga^{I}\gga^{J})_{AB}\,.
\ee
Simple computation with $\gamma$-matrices and \eqref{Kvec} gives
us the following final result for $K$
\be
K=\ff{1}{z}\gga_0(\gga_1+\gga_5)\,.
\ee
Clearly, $K^2=0$ while the radial scalar defined in \eqref{br} is
\be
r=\ff1z\,.
\ee
The explicit $sp(4)$ form of the planar subalgebra \eqref{plan}
that commutes with $K$ is also easy to find
\begin{align}
\quad T=&K=\ff12\gga_0(\gga_1+\gga_5)\,,\\
P_1=&\ff12\gga_2(\gga_1+\gga_5)\,,\\
\quad P_2=&\ff12\gga_3(\gga_1+\gga_5)\,,\\
L=&\ff12\gga_2\gga_3
\end{align}
with the commutation relations
\be
[T,P_{1,2}]=[T,L]=0\,,\qquad [P_{1},P_{2}]=0\,,\quad
[P_{1},L]=P_2\,,\quad [P_{2},L]=-P_1
\ee
that fulfill $u(1)\oplus iso(2)$ algebra.

\section*{Appendix B. Derivation of $sp(2)$ connection}

To proceed with the first order correction in the one-form sector
we have used flat connection $w_{\alpha \beta},\;
\bar{w}_{\dot{\alpha}\dot{\beta}}$ defined by \eqref{wflat}. It
was claimed to be flat which is synonymous to
\begin{equation}
\dr w_{\alpha \beta}-w_\alpha {}^\gamma  w_{\gamma \beta}=0.
\end{equation}
To check this fact one should use defining relations for $AdS_4$
geometry \eqref{adsOmega}, \eqref{adsTetrad} and differential
conditions on isometry parameters \eqref{param2a},
\eqref{param2b}. Differential of $k$ defined by \eqref{k} is given
by
\begin{multline}
\dr k_{\alpha \dot{ \alpha}}=2 \big(\mathbf{e}_{\sigma
\dot{\sigma}} k^{\sigma \dot{\sigma}}\big)k_{\alpha \dot{
\alpha}}+\omega_\alpha {}^\gamma k_{\gamma \dot{\alpha}}+2
\mathbf{e}_{\alpha \dot{\alpha}}+\frac{1}{r^2} \mathbf{e}_\beta
{}^{\dot{\gamma}} v_{\alpha \dot{\gamma}} v^{\beta}
{}_{\dot{\alpha}}+\bar{\omega}_{\dot{\alpha}} {}^{\dot{\gamma}}
k_{\alpha \dot{\gamma}}-\frac{1}{r^2} \varkappa_{\alpha} {}^\beta
\mathbf{e}_\beta {}^{\dot{\gamma}} \bar{\varkappa}_{\dot{\alpha}
\dot{\gamma}}.
\end{multline}
Now one is in a position to compute differential of $w_{\alpha
\beta}$ and product $w_\alpha {}^\gamma w_{\gamma \beta}$.
Straightforward computation by virtue of Fierz identity yields
\begin{multline}
\dr w_{\alpha \beta}=\omega_{\alpha} {}^\gamma  \omega_{\gamma \beta}-\half \omega_\alpha {}^\sigma \mathbf{e}_{\sigma \dot{\gamma}} k_\beta {}^{\dot{\gamma}}-\half \omega_\beta {}^\sigma \mathbf{e}_{\sigma \dot{\gamma}} k_{\alpha \dot{\gamma}}-\half \mathbf{e}_\alpha {}^{\dot{\gamma}} \omega_\beta {}^\sigma k_{\sigma \dot{\gamma}}-\half\mathbf{e}_\beta {}^{\dot{\gamma}} \omega_{\alpha} {}^\sigma k_{\sigma \dot{\gamma}}-\\
-\bar{\mathbf{H}}_{\dot{\gamma} \dot{\sigma}} k_\alpha
{}^{\dot{\sigma}} k_{\beta}
{}^{\dot{\gamma}}+\half\mathbf{H}_{\alpha
\beta}+\frac{1}{2r^2}\bar{\mathbf{H}}_{\dot{\gamma}\dot{\sigma}}
v_\beta {}^{\dot{\sigma}} v_\alpha
{}^{\dot{\gamma}}-\frac{1}{2r^2}
\bar{\mathbf{H}}_{\dot{\gamma}\dot{\sigma}} \varkappa_{\alpha
\beta} \bar{\varkappa}^{\dot{\gamma}\dot{\sigma}}
\end{multline}
and
\begin{multline}
w_\alpha {}^\gamma w_{\gamma \beta}=\omega_{\alpha} {}^\gamma \omega_{\gamma \beta}-\half \omega_\alpha {}^\gamma \mathbf{e}_{\gamma \dot{\gamma}} k_\beta {}^{\dot{\gamma}}-\half \omega_{\beta} {}^\sigma \mathbf{e}_{\sigma \dot{\gamma}} k_\alpha {}^{\dot{\gamma}}+\half\omega_{\alpha} {}^\sigma \mathbf{e}_\beta {}^{\dot{\gamma}} k_{\sigma \dot{\gamma}}-\half\mathbf{e}_\alpha {}^{\dot{\gamma}} \omega_\beta {}^\gamma k_{\gamma \dot{\gamma}}+\\
+\half \mathbf{H}_{\alpha \beta}-\half
\bar{\mathbf{H}}_{\dot{\gamma}\dot{\sigma}} k_\alpha
{}^{\dot{\gamma}} k_\beta {}^{\dot{\sigma}}.
\end{multline}
Here $\mathbf{H}_{\alpha \beta}$ and ${\bf \bar{H}}_{\dot{ \alpha}
\dot{ \beta}}$ are basic two-form defined as
\begin{equation}
\mathbf{H}_{\alpha \beta}= \mathbf{e}_{\alpha \dot{\gamma}}
\mathbf{e}_\beta {}^{\dot{\gamma}},\;\; \bar{\mathbf{H}}_{\dot{
\alpha} \dot{ \beta}}=\mathbf{e}_{\gamma \dot{ \alpha}}
\mathbf{e}^\gamma {}_{\dot{ \beta}}.
\end{equation}
After several simple cancellations one finds
\begin{equation}\label{w_flat1}
\dr w_{\alpha \beta}-w_{\alpha} {}^\gamma w_{\gamma \beta}=-\half
\bar{\mathbf{H}}_{\dot{\gamma}\dot{\sigma}} k_\alpha
{}^{\dot{\gamma}} k_\beta {}^{\dot{\sigma}}+\frac{1}{2r^2}
\bar{\mathbf{H}}_{\dot{\gamma} \dot{\sigma}} v_\beta
{}^{\dot{\sigma}} v_\alpha {}^{\dot{\gamma}}-\frac{1}{2 r^2}
\bar{\mathbf{H}}_{\dot{\gamma}\dot{\sigma}} \varkappa_{\alpha
\beta} \bar{\varkappa}^{\dot{\gamma}\dot{\sigma}}.
\end{equation}
Using definition of $k$, $r$ (given by \eqref{br}) and identity
\eqref{rel2} one can show that
\begin{multline}
k_\alpha {}^{\dot{\gamma}} k_\beta {}^{\dot{\sigma}}=\Big(-\frac{1}{r^2} \varkappa _\alpha {}^\sigma v_\sigma {}^{\dot{\gamma}} \Big)\Big(-\frac{1}{r^2} \varkappa_\beta {}^\xi v_\xi {}^{\dot{\sigma}}\Big)=-\frac{1}{r^4}\big(\varkappa _\alpha {}^\sigma v_\sigma {}^{\dot{\gamma}}\big)\big(v_\beta {}^{\dot{\xi}} \bar{\varkappa}_{\dot{\xi}} {}^{\dot{\sigma}}\big)=\\
=-\frac{1}{r^4}\Big[\big(v_\sigma {}^{\dot{\gamma}} v_\beta
{}^{\dot{\xi}}-v_\beta {}^{\dot{\gamma}} v_\sigma
{}^{\dot{\xi}}\big)\varkappa_\alpha {}^\sigma
\bar{\varkappa}_{\dot{\xi}} {}^{\dot{\sigma}}+v_\beta
{}^{\dot{\gamma}} v_\sigma {}^{\dot{\xi}} \varkappa_\alpha
{}^\sigma \bar{\varkappa}_{\dot{\xi}}
{}^{\dot{\sigma}}\Big]=-\frac{1}{r^4}\Big[ r^2 \epsilon_{\sigma
\beta} \bar{\epsilon}^{\dot{\gamma}\dot{\xi}}\varkappa_\alpha
{}^\sigma \bar{\varkappa}_{\dot{\xi}} {}^{\dot{\sigma}}+v_\beta
{}^{\dot{\gamma}} v_\sigma {}^{\dot{\xi}} \varkappa_\alpha
{}^\sigma \bar{\varkappa}_{\dot{\xi}} {}^{\dot{\sigma}}\Big].
\end{multline}
Previous identity allows one to bring \eqref{w_flat1} to the form
\begin{equation}
dw_{\alpha \beta}-w_{\alpha} {}^\gamma w_{\gamma
\beta}=\frac{1}{2r^4} \bar{\mathbf{H}}_{\dot{\gamma}\dot{\sigma}}
v_\beta {}^{\dot{\gamma}} v_\alpha {}^{\dot{\xi}} \varkappa_\alpha
{}^\sigma \bar{\varkappa}_{\dot{\xi}} {}^{\dot{\sigma}}
+\frac{1}{2r^2} \bar{\mathbf{H}}_{\dot{\gamma} \dot{\sigma}}
v_\beta {}^{\dot{\sigma}} v_\alpha {}^{\dot{\gamma}}.
\end{equation}
To simplify the remaining terms on the r.h.s. one should once
again apply identity \eqref{rel2}, hence one has
\begin{multline}
\dr w_{\alpha \beta}-w_{\alpha} {}^\gamma w_{\gamma \beta}=-\frac{1}{2r^4} \bar{\mathbf{H}}_{\dot{\gamma}\dot{\sigma}} v_\beta {}^{\dot{\gamma}} v_\alpha {}^{\dot{\rho}} \bar{\varkappa}_{\dot{\rho}} {}^{\dot{\xi}} \bar{\varkappa}_{\dot{\xi}} {}^{\dot{\sigma}}+\frac{1}{2r^2} \bar{\mathbf{H}}_{\dot{\gamma} \dot{\sigma}} v_\beta {}^{\dot{\sigma}} v_\alpha {}^{\dot{\gamma}}=\\
=-\frac{1}{2r^4}\bar{\mathbf{H}}_{\dot{\gamma}\dot{\sigma}}
v_\beta {}^{\dot{\gamma}} v_\alpha {}^{\dot{\rho}}r^2
\delta_{\dot{\rho}}^{\dot{\sigma}}+\frac{1}{2r^2}
\bar{\mathbf{H}}_{\dot{\gamma} \dot{\sigma}} v_\beta
{}^{\dot{\sigma}} v_\alpha {}^{\dot{\gamma}}=0.
\end{multline}
Flatness of $\bar{w}_{\dot{ \alpha}\dot{\beta}}$ can be checked
analogously.

\section*{Appendix C. Checking the adjoint module}
In solving \eqref{weq} the two conditions should be satisfied
separately. Indeed, being a differential two-form, \eqref{weq}
splits into the part proportional to ${\bf E}^{\al\gga}{\bf
E}_{\gga}{}^{\gb}$ and to ${\bf E}^{\al\gb}{\bf E}$. Each of these
should be equal to zero. We have checked that \eqref{anz} passes
through ${\bf E}^{\al\gga}{\bf E}_{\gga}{}^{\gb}$ -- part and
fixes the form of solution \eqref{wsol} unambiguously. It is
therefore win or lose for the remaining ${\bf E}^{\al\gb}{\bf E}$.
Let us check here that \eqref{wsol} satisfies it too.

Sector ${\bf E}^{\al\gb}{\bf E}$ is contributed by the first and
third terms of \eqref{weq}. Consider those in detail. The first
term is
\begin{align}\label{eq2}
&\nabla\left({\bf
E}^{\gb\gb}\int_{0}^{1}\dr\tau\p_{\gb}\p_{\gb}f(\ff{1}{2r^2}\gk_{\al\al}(\tau
y+(1-\tau)\bar y)^{\al(2)})\right)=\\
&=-{\bf E}^{\gb\gb}{\bf E}\,\p_{\gb}\p_{\gb} \int_{0}^{1}\dr\tau
\ff{f'}{2r}\gk_{\al\al}(\tau y+(1-\tau)\bar y)^{\al(2)}\,,\nn
\end{align}
where $f'(x)=\ff{\dr}{\dr x}f(x)$. Let us now note that for
$f(\tau y+(1-\tau)\bar y)$ the differentiation over the argument
can be rewritten as
\be
\p f=\left(\ff{\p}{\p y}+\ff{\p}{\p\bar y}\right) f
\ee
and thus \eqref{eq2} boils down to
\be
-{\bf E}^{\gb\gb}{\bf E} \left(\ff{\p}{\p y}+\ff{\p}{\p\bar
y}\right)_{\gb(2)}^2\int_{0}^{1}\dr\tau
\ff{f'}{2r}\gk_{\al\al}(\tau y+(1-\tau)\bar y)^{\al(2)}\,.
\ee
To calculate the third term in \eqref{weq} we use that
\be
\left(y^{\al}\ff{\p}{\p\bar y^{\al}}+\bar y^{\al}\ff{\p}{\p
y^{\al}}\right) \left(\ff{\p}{\p y}+\ff{\p}{\p\bar
y}\right)_{\gb(2)}^2=\left(\ff{\p}{\p y}+\ff{\p}{\p\bar
y}\right)_{\gb(2)}^2 \left(y^{\al}\ff{\p}{\p\bar y^{\al}}+\bar
y^{\al}\ff{\p}{\p y^{\al}}\right)-2 \left(\ff{\p}{\p
y}+\ff{\p}{\p\bar y}\right)_{\gb(2)}^2\,.
\ee
Now, evaluating
\begin{align}
\left(y^{\al}\ff{\p}{\p\bar y^{\al}}+\bar y^{\al}\ff{\p}{\p
y^{\al}}\right) f=\ff{f'}{r^2}\gk_{\al\al}(\tau y+(1-\tau)\bar
y)^{\al}(\tau\bar y+(1-\tau)y)^{\al}
\end{align}
and using that
\be
\ff{f'}{r^2}\gk_{\al\al}(\tau y+(1-\tau)\bar y)^{\al}(y-\bar
y)^{\al}=\ff{\p}{\p\tau} f
\ee
we have
\begin{align}
&\left[\nabla-\ff{r}{2}{\bf E}\left(y^{\al}\ff{\p}{\p\bar
y^{\al}}+\bar y^{\al}\ff{\p}{\p y^{\al}}\right) \right] {\bf
E}^{\gb\gb}\int_{0}^{1}\dr\tau \p_{\gb}\p_{\gb} f(\tau
y+(1-\tau)\bar y)=\\
&=\ff{r}{2}{\bf E}^{\gb\gb}{\bf E} \int_{0}^{1}\dr\tau
(1-2\tau)\ff{\p}{\p\tau} \p_{\gb}\p_{\gb} f(\tau y+(1-\tau)\bar
y)-r{\bf E}^{\gb\gb}{\bf E} \int_{0}^{1}\dr\tau \p_{\gb}\p_{\gb}
f(\tau y+(1-\tau)\bar y)\,.\nn
\end{align}
Integrating by parts we finally obtain
\begin{align}
&\left[\nabla-\ff{r}{2}{\bf E}\left(y^{\al}\ff{\p}{\p\bar
y^{\al}}+\bar y^{\al}\ff{\p}{\p y^{\al}}\right) \right] {\bf
E}^{\gb\gb}\int_{0}^{1}\dr\tau \p_{\gb}\p_{\gb} f(\tau
y+(1-\tau)\bar y)=\\
&=-\ff{r}{2}{\bf E}^{\gb\gb}{\bf
E}(\p_{\gb}\p_{\gb}f(y)+\bar\p_{\gb}\bar\p_{\gb}f(\bar y))\nn
\end{align}
and so we conclude that \eqref{wsol} is the solution.



\end{document}